\documentclass[A4paper,11pt]{article}

\pdfoutput=1
\usepackage{latexsym}
\usepackage{epsfig}
\usepackage{verbatim}
\usepackage{shadow}
\usepackage{amssymb}
\usepackage{amsmath}
\usepackage{graphicx}

\newtheorem{theorem}{\sc Theorem}[section]
\newtheorem{lemma}[theorem]{\sc Lemma}
\newtheorem{corollary}[theorem]{\sc Corollary}
\newtheorem{fact}[theorem]{\sc Fact}
\newtheorem{remark}[theorem]{\sc Remark}

\newcommand{\eps}{\varepsilon}

\newcommand{\proofend}{{\medskip\medskip}}
\newcommand{\proof}{{\noindent\em Proof. }}

\author{
  {\sc Bernard Chazelle}
\thanks{Department of Computer Science,
       Princeton University, 
{\tt chazelle}@{\tt cs.princeton.edu }}
}

\title{
The Total $s$-Energy of a Multiagent System
\thanks{A preliminary version of this work
appeared in {\em Proc. 26th Annual ACM
Symposium on Computational Geometry}, 2010.
This work was supported in part by NSF grants CCF-0634958,
CCF-0832797, CCF-0963825, and CCF-1016250.
}}

\vspace{4cm}

\date{}

\begin{document} \maketitle

\begin{abstract}
We introduce the {\em total $s$-energy} of a
multiagent system with time-dependent links. 
This provides a new analytical lens
on bidirectional agreement dynamics, which we use
to bound the convergence rates of dynamical systems
for synchronization, flocking,
opinion dynamics, and social epistemology.
\end{abstract}

\vspace{4cm}

\section{Introduction}\label{introduction}

We introduce an analytical device
for the study of multiagent agreement systems.
Consider an infinite sequence of graphs,
$G_0,G_1,G_2, \ldots$, each one defined on $n$ nodes labeled
$1,\ldots, n$. We assume that each graph $G_t$ is embedded
in Euclidean $d$-space and we let $x_i(t)\in {\mathbb R}^d$
denote the position of node $i$ at time $t$.
The {\em total $s$-energy} $E(s)$ of the embedded graph sequence 
encodes all of the edge lengths:
\begin{equation}\label{E(s)}
E(s)= \sum_{t\geq 0}\,
\sum_{(i,j)\in G_t} \|x_i(t)-x_j(t)\|_2^{s} \, ,
\end{equation}
where the exponent $s$ is a real (or complex) variable.
The definition generalizes both the Dirichlet form derived from the 
graph Laplacian and 
the Riesz $s$-energy of points on a sphere. Sometimes, variants
of the total $s$-energy 
are more convenient; for example, we will use
the {\em kinetic $s$-energy},
\begin{equation}\label{K(s)}
K(s)= \sum_{t\geq 0}\,
\sum_{i=1}^n \|x_i(t+1)-x_i(t)\|_2^{s} \, .
\end{equation}
Note that these definitions make no assumptions about
the {\em temporal network}, which is the name 
given to a graph sequence sharing the same node set.
There is no reason to think that $s$-energies should even be finite,
let alone useful: for example,
$E(0)$ is usually infinite. In fact, it is immediate to embed 
a temporal network so as to make its
total $s$-energy diverge everywhere, so one clearly needs
assumptions on the embeddings. In this paper we consider
the case of {\em multiagent 
agreement systems}~\cite{Moreau2005}, which we define 
in the next section. There are two kinds: bidirectional
and nonbidirectional. We consider only the former type
in this work. We thus assume that each $G_t$ is undirected,
meaning that if $(i,j)$ is an edge then so is $(j,i)$.
(The directed case is quite different and warrants
a separate treatment.) 

We use the total $s$-energy to
bound the convergence rates of classical systems for
opinion dynamics~(\S\ref{opinion-dynamics}),
social epistemology~(\S\ref{truthseeksys}),
Kuramoto synchronization~(\S\ref{sync}), and
bird flocking~(\S\ref{flocking}).
We deal only with discrete-time dynamical systems or, 
as in~\cite{Moreau2005}, time-1 maps
of continuous systems. We also improve
a classic bound for the products of stochastic matrices~(\S\ref{prodstoch}).

Our proofs are algorithmic and make no use of 
algebraic graph theory (with a single exception for reversible systems
where we turn to a standard $\ell_2$ argument~\S\ref{multiagent-dynamics}).
In particular, the proofs focus on the agents rather than 
on the matrices governing their dynamics. In fact, the proofs
themselves can be viewed as dynamical systems which
embed convergence measures directly within 
the inter-agent communication. We hope this perhaps 
opaque comment finds clarification below and that the benefits of 
an algorithmic approach to multiagent dynamics
becomes apparent~\cite{chazFlockSODA, chazICS}.

\subsection{Multiagent Dynamics}\label{multiagent-dynamics}

Moreau~\cite{Moreau2005} introduced a geometric
framework for multiagent agreement dynamics of appealing generality.
He established convergence criteria based on connectivity
in both the directed case and the undirected one.
We seek to analyze the dynamics of bidirectional 
systems {\em without any connectivity assumptions}:
specifically, our goal is to provide bounds on convergence rates that hold
in {\em all} cases.

\smallskip
\paragraph{Bidirectional agreement systems.}

The one-dimensional case features all of the ideas
and it is straightforward to extend our analysis to $d>1$;
we briefly mention how to do that below. 
For simplicity, therefore, we assume that $d=1$.
The model involves $n$ agents located at the points 
$x_1(t),\ldots, x_n(t)$ in ${\mathbb R}$
at any time $t\geq 0$. The input consists of
their positions at time $t=0$, together with an infinite
sequence $(G_t)_{t\geq 0}$ of undirected 
graphs over $n\geq 2$ nodes (the agents);
each node has a self-loop. 
These graphs represent the various configurations
of a communication network changing over time.
The sequence need not be 
known ahead of time: in practice, the system will often be embedded
in a closed loop and the next $G_{t}$ will be a function of
the configuration at time $0,\ldots, t-1$. The strength
of the model is that it makes no assumption about the generation
of the temporal networks nor about their connectivity properties. 
In the case of directed graphs,
such a level of generality precludes blanket statements
about convergence; bidirectionality, on the other hand, allows
such statements. The neighbors of $i$
form the set $N_i(t)= \{\,j\,|\, (i,j)\in G_t\,\}$,
which includes $i$.
At time $t$, each agent $i$ moves {\em anywhere} within the 
interval formed by its neighbors, though not too close
to the boundary: formally, if $m_{i,t}$ is the minimum of
$\{\,x_j(t)\,|\, j\in N_i(t)\,\}$ and $M_{i,t}$ is the maximum,
then 
\begin{equation}\label{gen-dyn}
(1-\rho)m_{i,t} + \rho M_{i,t}\leq x_i(t+1)\leq 
\rho m_{i,t} + (1-\rho) M_{i,t} , 
\end{equation}
where $0<\rho\leq 1/2$ is the (time-independent) {\em agreement parameter},
fixed once and for all. 
All the agents are updated in parallel at 
each step $t=0,1,2,$ etc.
We conclude the presentation of bidirectional agreement systems
with a few remarks.

\begin{figure}[htb]
\begin{center}
\hspace{0cm}
\includegraphics[width=5cm]{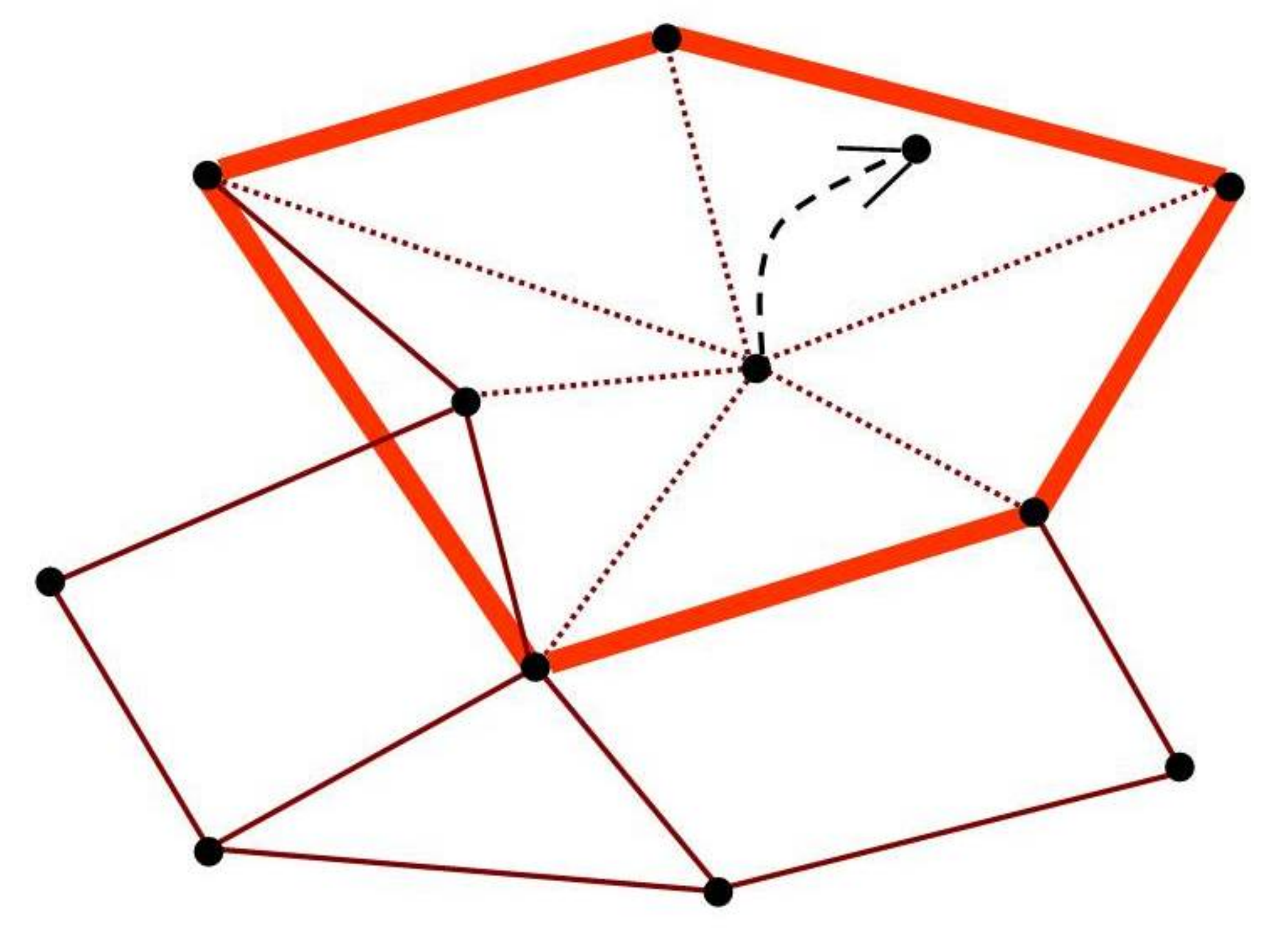}
\end{center}
\vspace{-0.2cm}
\caption{\small 
The agent can move anywhere inside the pentagon but
may not touch the thick boundary.
\label{fig-rules}}
\end{figure}

\begin{itemize}
\item
The model describes a {\em nondeterministic} dynamical system.
This refers to the fact that
the sequence of graphs, as well as the particular motion of the agents,
are left completely arbitrary within the constraints imposed
by~(\ref{gen-dyn}): they could be decided ahead of time or, as is more 
common, endogenously in a closed-loop system; we give several
examples below. The embedding at time 0 is provided as input and, from then on,
all subsequent embeddings are generated by the system itself in abidance by
rule~(\ref{gen-dyn}). It may sound surprising at first that
one can prove convergence in the presence of such high nondeterminism
and without the slighest assumption about connectivity.

\item
Bidirectionality does not imply symmetry among neighbors.
In fact, the behavior of
neighboring agents may be completely different.
The condition $\rho>0$ is essential.
Without it, a two-agent system with a single edge
could see the agents swap places forever without ever converging.
This simple example shows that one may legally move the two agents
toward each other so that their distance decreases by a factor
of merely $1-2\rho$ at each step. This shows that no worst-case
convergence rate can be faster than $e^{-2\rho t}$.

\item
There are several ways to extend the model to higher dimension.
Perhaps the easiest one is to assume that agent $i$ is positioned
at $x_i(t)\in {\mathbb R}^d$ and then enforce~(\ref{gen-dyn}) along each
dimension. This is equivalent to having $d$ one-dimensional systems
sharing the same temporal network; it is the method we use
in this paper. A different, coordinate-free approach stipulates
that agent $i$ may move anywhere within the convex hull
$C_i(t)$ of its neighbors $\{\, x_j(t)\,|\, j\in N_i(t)\,\}$, 
but not too close to the boundary (Figure~\ref{fig-rules}).
This requires shrinking 
$C_i(t)$ by a factor of $1-\rho$ centrally toward a well-chosen
center: for example, the {\em L\"owner-John center} of $C_i(t)$,
which is uniquely defined as the center of the minimum-volume ellipsoid 
that encloses $C_i(t)$~\cite{grotschelLS}.
\end{itemize}

\vspace{0.5cm}

\bigskip

Much of the previous work on agreement systems 
has been concerned with  
conditions for consensus (ie, for all agents to come together), 
beginning with the pioneering work of~\cite{tsitsiklis84,tsitsiklisBA}
and then~\cite{angeliB, blondelHOT05, caoAM, caoSM, HendrickxB, 
jadbabaieLM03, liwang2004, Moreau2005, olshevskyT-06}.
Bounds on the convergence rate have been
obtained under various connectivity 
assumptions~\cite{caoSM, olshevskyT-06} and
for specialized closed-loop systems~\cite{chazFlockPaper, martinezBCF07}.
The convergence of bidirectional agreement systems
can be derived from the techniques
in~\cite{HendrickxB,lorenz05,Moreau2005}.
Bounding the convergence rate, however,
has been left open. This is the main focus of this paper.
Before stating our results in the next section,
we discuss a few extensions of the model.

\smallskip\smallskip
\paragraph{The fixed-agent agreement model.}

\begin{figure}[htb]
\begin{center}
\hspace{.2cm}
\includegraphics[width=6cm]{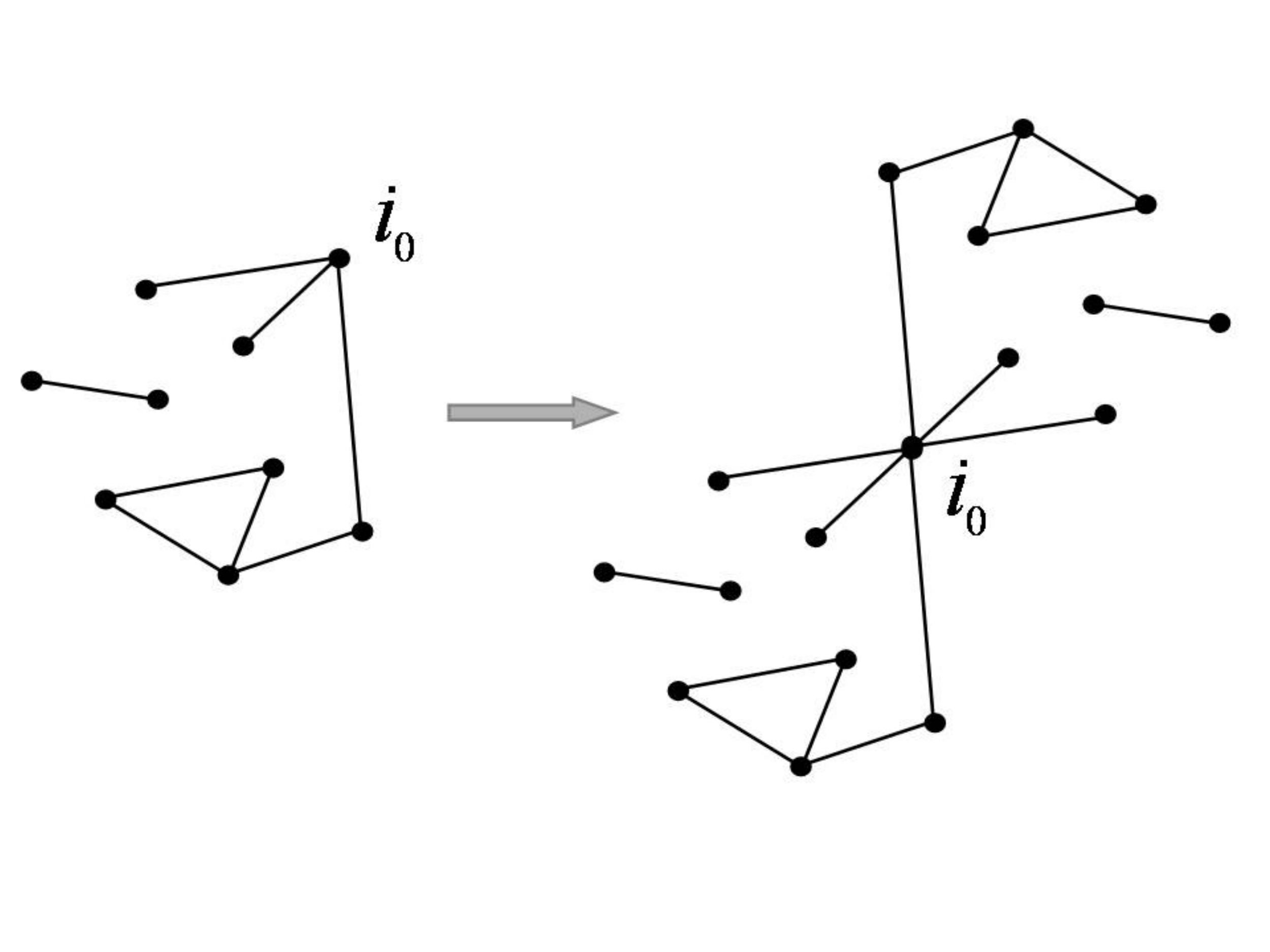}
\end{center}
\vspace{-0.2cm}
\caption{\small Reflecting the system
about the agent $i_0$ that we wish to fix.
\label{fig-reflection}}
\end{figure}

We can fix one agent if we so desire.
By this, we mean skipping the update
rule at an arbitrary agent $i_0$, selected ahead of time---or 
equivalently, directing all edges incident to $i_0$ toward that node. 
To see why, create the point reflection
of the $n-1$ mobile agents about $i_0$
to create a bidirectional system of $2n-1$ agents.
Figure~\ref{fig-reflection} illustrates this process
in two dimensions for visual clarity. 
We duplicate each graph $G_t$, with the exception
of the fixed agent $i_0$. In this way, at time $t$, each edge $(i,j)$ is given
a duplicate $(i',j')$. Placing the origin of a Cartesian
coordinate system
at $x_{i_0}(0)$, we position agent $i'$ at time $0$
so that $x_{i'}(0)= - x_i(0)$, which inductively implies
that $x_{i'}(t)= - x_i(t)$ for all $t\geq 0$. No edges connect
the two copies of the original graphs.
Every mobile agent (and its reflected copy) mimics
the behavior of its counterpart in the original $n$-agent system
while respecting~(\ref{gen-dyn}). 
The fixed agent always lies at the midpoint of the smallest interval
enclosing its neighbors; therefore, it does not need to move,
even for the maximum value of $\rho$ allowed, which is $1/2$.
To summarize, any $n$-agent agreement system 
with one fixed agent can be simulated with
a $(2n-1)$-agent bidirectional agreement system with 
the same value of $\rho$ and at most twice the diameter.
We apply this result to truth-seeking systems in~\S\ref{truthseeksys}.

\vspace{0.1cm}
\smallskip

\smallskip
\paragraph{Reversible agreement systems.}

Assign to each agent $i$ a time-independent 
{\em motion parameter} $q_i\geq |N_i(t)|$ 
and define the mass center of the agent's neighbors as
$$\mu_i(t)= \frac{1}{|N_i(t)|} \sum_{j\in N_i(t)} x_j(t)\, .$$
A {\em reversible agreement system} satisfies the transition:
$$
x_i(t+1)= x_i(t) + \frac{|N_i(t)|}{q_i} \Bigl( \mu_i(t) - x_i(t) \Bigr).
$$
The agents obey the dynamics $x(t+1)= P(t)x(t)$, where
\begin{equation}\label{reversible-pij}
p_{ij}(t)=
\begin{cases}
\, 1 - ( |N_i(t)|-1 )/q_i &\text{ if $i=j$; } \\
\, 1/q_i &\text{ if $i\neq j\in N_i(t)$; } \\
\, 0 &\text{ else.}
\end{cases}
\end{equation}
A quick examination of~(\ref{gen-dyn}) shows that
the dynamics, indeed,
defines an agreement system with parameter $\rho= 1/\max_i q_i$.
Why reversible? We take note of the identity
$q_ip_{ij}(t)= q_jp_{ji}(t)$. This is the standard
balanced condition of a reversible Markov chain, with
$(q_i)$ in the role of the stationary distribution (up to scaling).
Indeed, we easily verify that
the sum $\sum_i q_i x_i(t)$ is independent of $t$
and that a lazy random walk in a graph is a special case
of a reversible agreement system.
The latter is much more general, of course, since
the graph can change over time.
We note that, if each node has its own degree fixed over time,
then moving each agent to the mass center of its neighbors
satisfies reversibility and hence, as we shall see, fast convergence.
This is equivalent to setting $q_i=|N_i(t)|$.

The definition of a reversible system is simple yet somewhat
contrived. Does the concept lend itself to intuition? Yes.
At each step, each agent picks a set of neighbors (the graph $G_t)$
and slides toward their mass center---but not all the way!
The agent might have to stop before hitting the mass center.
When? This is where the $q_i$'s come in.
They ensure that 
the $q_i$-weighted mass center of the whole system stays put.
Not to have that mass center wiggle around
and produce exponentially small coefficients
is a key reason why
reversibility implies faster convergence.
We flesh out this intuition in~\S\ref{sec-convergence}.

\smallskip

\paragraph{The matrix approach.}

It is customary to model agreement systems 
by using products of stochastic matrices:
$x(t+1)= P(t)x(t)$, where 
$x(t)=(x_1(t),\ldots, x_n(t))^T$
and $P(t)$ is a row-stochastic matrix whose
entries $p_{ij}(t)$ are positive for all $i,j$ with $j\in N_i(t)$. 
Bidirectionality means
that $p_{ij}(t)$ and $p_{ji}(t)$ should be both positive or both zero,
which is a form of {\em mutual confidence}~\cite{lorenz05}.
Typically, one also requires a uniform lower bound on {\em each} nonzero entry
of the matrices. We observe that condition~(\ref{gen-dyn}) is
not nearly as demanding: all we require is that, if the agent
$i$ has at least one neighbor (besides itself), then the entries corresponding
to the leftmost and rightmost neighbors $l(i)$ and $r(i)$ 
should be at most $1-\rho$. 
These conditions have a natural interpretation that we summarize below:
for all~$t$,
\begin{equation}\label{pij-rho}
\begin{cases}
\, \text{{\em Mutual confidence}: \ No pair $p_{ij}(t),p_{ji}(t)$ has exactly 
one zero}; \\
\, \text{{\em No extreme influence}: \ For any nonisolated agent $i$,
$\max\{p_{il(i)}(t), p_{ir(i)}(t)\} \leq 1-\rho$.}
\end{cases}
\end{equation}
Conditions~(\ref{pij-rho}) are weaker than the usual set of
three constraints associated 
with the bidirectional case~\cite{HendrickxB,lorenz05},
which, besides mutual confidence, includes:
{\em self-confidence} (nonzero diagonal entries)
and {\em nonvanishing confidence} (lower bound on all nonzero entries).
Our model requires bounds on only two entries per matrix row.
Previous work~\cite{blondelHOT05, chazFlockSODA, lorenz05}
highlighted the importance of self-confidence $(p_{ii}(t)>0$)
for the convergence of agreement systems.
Our results refine this picture:
{\em To reach harmony in a group with bidirectional
communication, individuals may be
influenced extremely by
non-extreme positions but must be influenced
non-extremely by extreme positions ($m_{i,t}$ or $M_{i,t}$).}
In the case of a two-agent system,
this maxim coincides with the need for self-confidence;
in general, the latter is not needed.
We conclude this comment about the matrix representation
of agreement systems by emphasizing that the total $s$-energy
seeks to move the focus away from the matrices themselves
and, instead, reason about the agents' motion in phase space and their 
temporal communication network.

\smallskip

\paragraph{Random walks and ergodicity.}

At the risk of oversimplifying, one might say that
to understand agreement systems is to undestand
backward products of stochastic matrices, 
$$P(t)P(t-1)\cdots P(1)P(0),$$
as $t$ grows to infinity.
Forward products $P(0)P(1)\cdots P(t)$, for $t\rightarrow \infty$,
are different but much can be inferred
about them from the backward variety.
A forward product of stochastic matrices models 
a random walk in a temporal network: imagine walking randomly
in a graph that may change at every step. These 
have been studied by computational complexity theorists,
who call them {\em colored random walks}~\cite{condonhernek94, CondonL}. 
This connection suggests that a complete theory of 
agreement systems would need to include, as a special case,
a theory of discrete-time Markov chains. As we shall see,
the total $s$-energy allows us to retrieve classical
mixing bounds for random walks in undirected graphs.

\vspace{0.5cm}
\begin{figure}[htb]
\begin{center}
\hspace{.2cm}
\includegraphics[width=8cm]{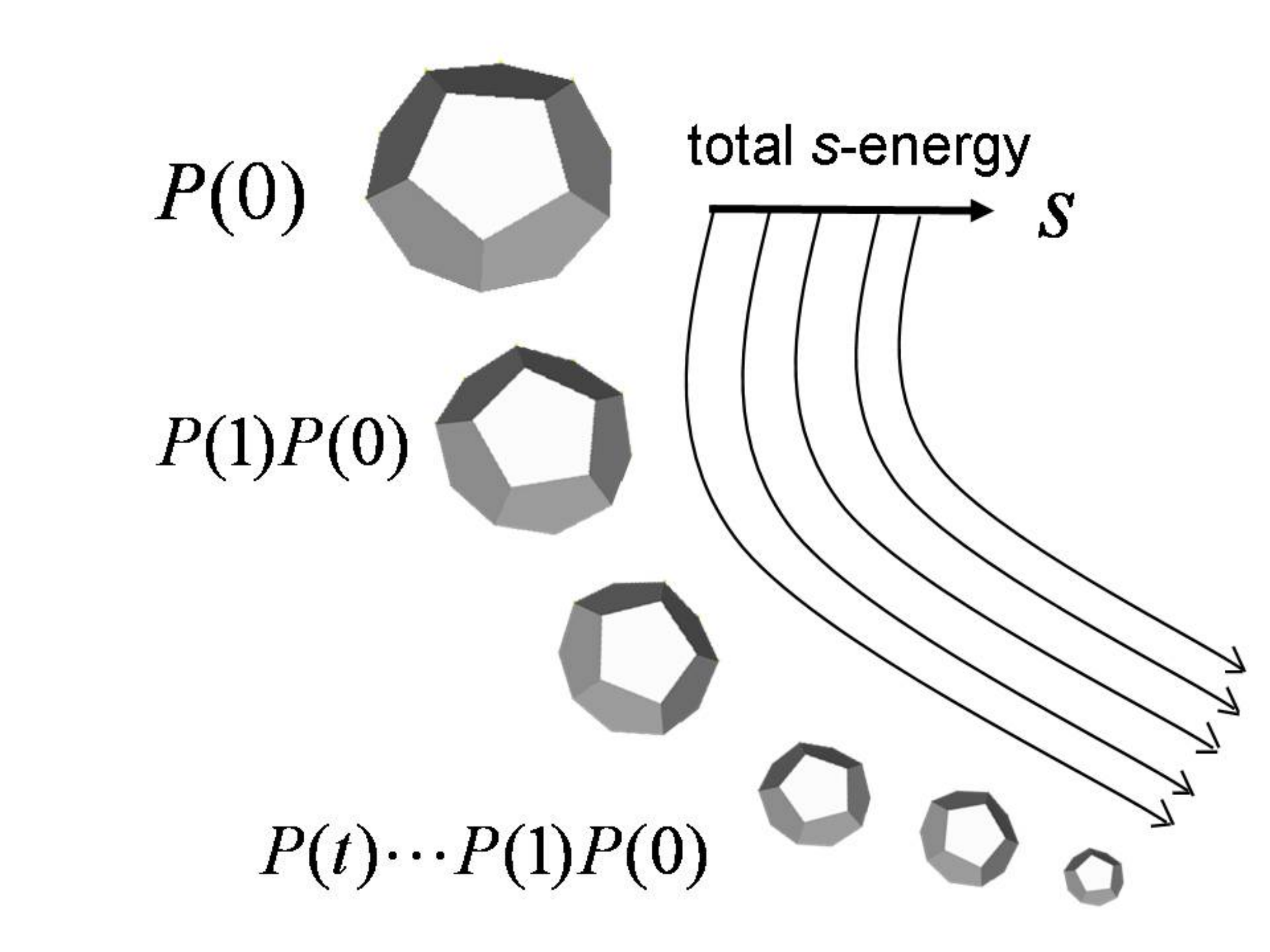}
\end{center}
\caption{\small The total $s$-energy is a global
instrument to track how fast the polytopes shrink.
\label{globaltracker}}
\vspace{0.3cm}
\end{figure}

A general principle behind the convergence of products
of stochastic matrices is that, if all goes well,
as $t$ grows, the product will
tend to a matrix of rank one or a (possibly permuted) block-diagonal matrix
with blocks of rank one. Many analytical devices have
been designed to keep track of this evolution, most of which
fall into the category of {\em ergodicity coefficients}~\cite{seneta06}.
There is a simple geometric interpretation of this which is worth
a short detour. From a stochastic matrix such as $P(0)$, construct
a convex polytope by taking the convex hull, denoted $\text{\tt conv}P(0)$,
of the points formed by the rows of $P(0)$: here, each row
forms the $n$ coordinates of a point.  When we multiply
$P(1)$ by $P(0)$, each row of the product is a convex combination
of the rows of $P(0)$, so the corresponding point lies inside
the convex hull $\text{\tt conv}P(0)$; therefore
$\text{\tt conv}\{P(1)P(0)\}\subseteq \text{\tt conv}P(0)$.
A coefficient of ergodicity is typically a measure of how quickly
the nesting shrinks: it might keep track of the width, diameter, volume,
or any other ``shrinking'' indicator:
$$\text{\tt conv}\{P(t)\cdots P(0)\}\subseteq
  \text{\tt conv}\{P(t-1)\cdots P(0)\}\subseteq
  \cdots \subseteq \text{\tt conv}\{P(1)P(0)\}\subseteq \text{\tt conv}P(0) .$$
For example, when all the matrices are identical (as in a Markov chain),
the spectral gap gives us an $\ell_2$-norm tracker of the contraction.
What all coefficients of ergodicity have in common is that they
are {\em local} instruments: they measure the contraction
from one step to the next. The total $s$-energy, instead,
is a {\em global} instrument. It monitors the shrinking over all time steps
in a global fashion: the parameter $s$ plays the role of
frequency in Fourier analysis and allows us to choose the
correct ``frequency'' at which we want to monitor the shrinking process.
This gives us Chernoff-like bounds on the distribution of the edge lengths.

\subsection{The Total $s$-Energy}\label{thetotalsenergy}

There is no obvious reason why the total $s$-energy, as defined
in~(\ref{E(s)}), should {\em ever} converge,
so we treat it as a formal series for the time being.
We prove that it converges for any real $s>0$ 
and we bound its maximum value,
$E_n(s)$, over all moves and $n$-node graph sequences.
We may assume that all the agents start out in the unit interval
(which, of course, implies that they remain there at all times).
The justification is that 
the total $s$-energy obeys a power-law under
scaling: $x\mapsto Cx$ implies that $E_n(s)\mapsto C^s E_n(s)$.
We also assume throughout the remainder
of this paper that $\rho$ is smaller than a suitable constant.
All the proofs of the results below are deferred to~\S\ref{Proofs}.

\begin{theorem}\label{En(s)bound}
$\!\!\! .\,\,$
The maximal total $s$-energy of an $n$-agent 
bidirectional agreement system with unit initial diameter
satisfies

\begin{equation*}
E_n(s)\leq \begin{cases}
\, \rho^{-O(n)} &  \text{ for $\, s=1$};  \\
\,  s^{1-n}\rho^{-n^2 -O(1)} & \text{ for $\, 0<s<1$}.
\end{cases}
\end{equation*}
There is a lower bound of 
$O(\rho)^{-\lfloor n/2\rfloor}$ on $E_n(1)$ 
and of $s^{1-n} \rho^{-\Omega(n)}$
on $E_n(s)$ for $n$ large enough, any
$s\leq s_0$, and any fixed $s_0<1$.
\end{theorem}

\smallskip\smallskip

The asymptotic notation hides the presence of absolute constant factors.
For example, $\rho^{-O(n)}$, $O(\rho)^{-\lfloor n/2\rfloor}$,
and $\rho^{-\Omega(n)}$ mean, respectively, at most 
$\rho^{-an}$, at least $(b\rho)^{-\lfloor n/2\rfloor}$,
and at least $\rho^{-cn}$, for some suitable constants $a,b,c>0$.
Since no edge length exceeds 1, 
$E_n(s)\leq E_n(1)$ for $s\geq 1$, and so the theorem proves
the convergence of the total $s$-energy for all $s>0$.

When the temporal network always stays connected, 
it is useful to redefine the total
$s$-energy as the sum of the $s$-th powers
of the diameters. Its maximum value, for unit initial diameter,
is denoted by 
$$E^D_n(s)=
\sum_{t\geq 0}\, \Bigl( \text{diam}\, \{x_1(t),\ldots, x_n(t)\} \Bigr)^s.$$
In dimension $d=1$, the diameter is the length of the smallest
enclosing interval. The following result is the sole breach
of our pledge to avoid any connectivity assumption.

\smallskip

\begin{theorem}\label{En(s)reversible}
$\!\!\! .\,\,$
The maximal diameter-based total $s$-energy of 
a connected $n$-agent reversible agreement system 
with unit initial diameter satisfies
$$
n^{-2} E_n(s)\leq 
E^D_n(s)\leq 
\frac{2n}{s}\Bigl( \frac{2n}{\rho} \Bigr)^{s/2+1},
$$
for all $0<s\leq 1$.
\end{theorem}

\smallskip

\vspace{0.5cm}
\begin{figure}[htb]
\begin{center}
\hspace{.2cm}
\includegraphics[width=8cm]{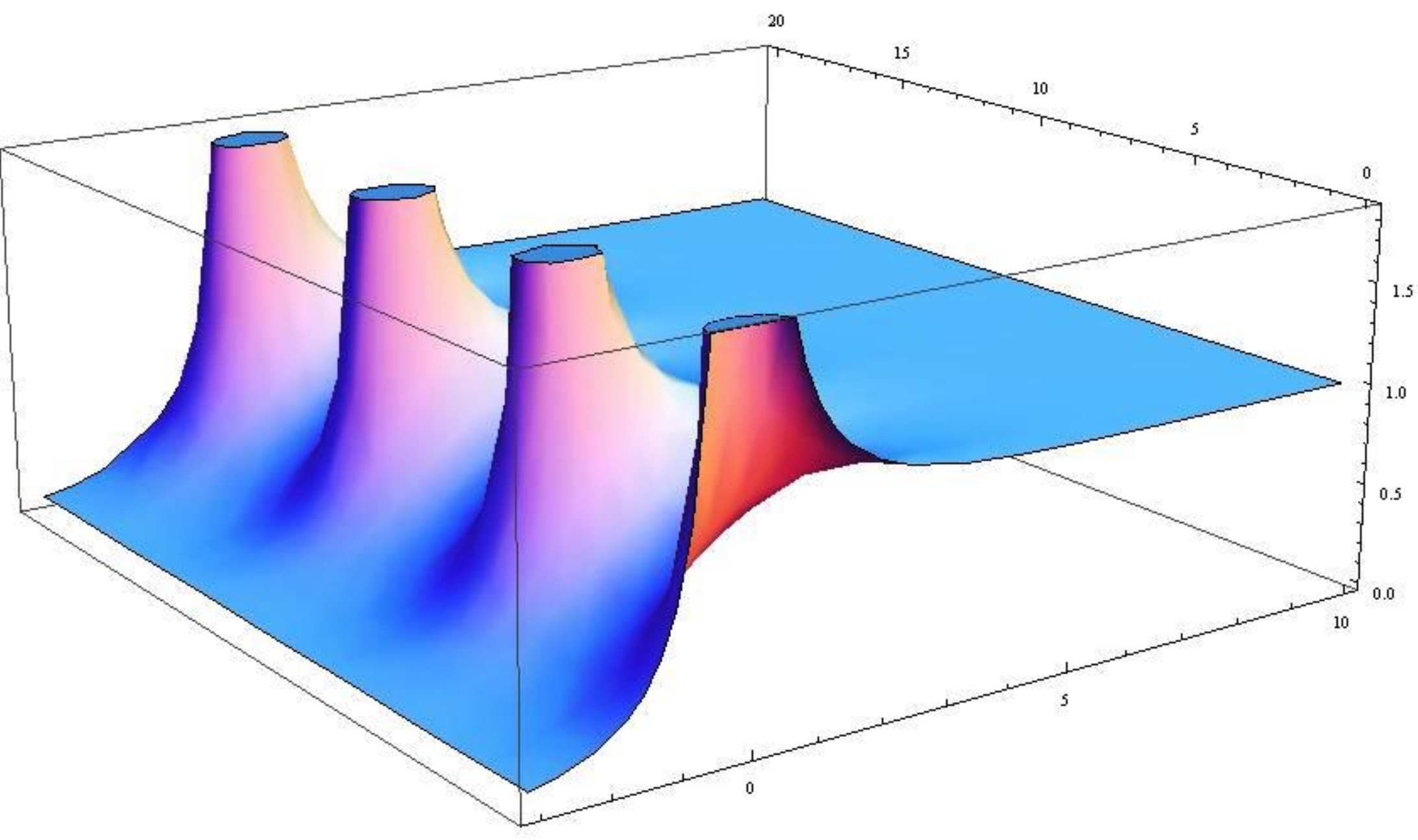}
\end{center}
\caption{\small By analytic continuation,
the maximum total $s$-energy of a two-agent
system is a meromorphic function over the whole
complex plane; the function is depicted in absolute value,
with the real axis stretching to the right.
\label{2agent-continuation}}
\vspace{0.3cm}
\end{figure}

We proceed with general remarks about the function $E(s)$.
All of the terms in the series are nonnegative,
so we can assume them rearranged
in nonincreasing order. This allows us to express the 
total $s$-energy as a general Dirichlet series:
\begin{equation}\label{dirichlet-series}
E(s)= \sum_{k\geq 1} n_k e^{-\lambda_k s},
\end{equation}
where $\lambda_k= -\ln d_k$ and
$n_k$ is the number of edges of length $d_k$.  
Thus, $E(s)$ is the Laplace transform
of a sum of scaled Dirac delta functions centered at $x=\lambda_k$.
This implies that the total $s$-energy can be inverted
and, hence, provides a lossless encoding of 
the edge lengths.
We show that $E(s)$ converges for any real $s>0$.
By the theory of Dirichlet series~\cite{hardyR},
it follows that $E(s)$ is uniformly convergent
over any finite region
${\mathcal D}$ of the complex plane within $\Re (s)\geq r$,
for any $r>0$; furthermore,
the series defines an analytic function over ${\mathcal D}$.
It is immediate to determine the maximum $s$-energy
of a 2-agent system with unit initial diameter. For $\rho=1/2-1/2e$,
$E_2(s)= \sum_t (1-2\rho)^{st}= 1/(1-e^{-s})$; therefore,
writing $s=x+iy$, it satisfies (Figure~\ref{2agent-continuation}):
$$
|E_2(s)|=1/\sqrt{ 1-2e^{-x}\cos y + e^{-2x} } \, .
$$
The singularities are the simple poles $s=2\pi ik$, for all $k$.
The maximal total $s$-energy can be continued meromorphically
over the whole complex plane. Note that this
is obviously false for {\em nonmaximal} $s$-energies:
for example, the function $\sum_k e^{-sk!}$ is a valid total
$s$-energy, but its singularities form a dense
subset of its line of convergence (the imaginary axis),
hence an impassable barrier for
any analytic continuation into $\Re (s)< 0$.

\vspace{0.5cm}
\begin{figure}[htb]
\begin{center}
\hspace{0cm}
\includegraphics[width=8cm]{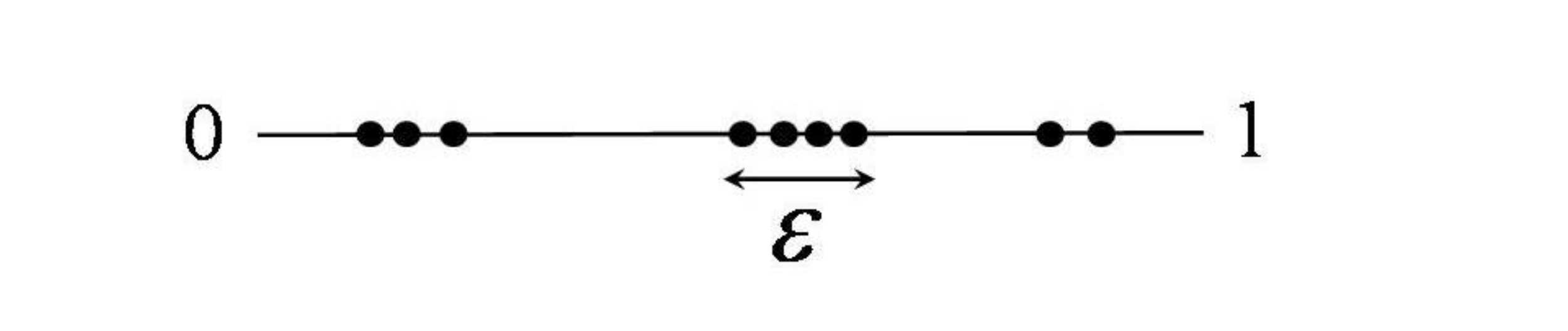}
\end{center}
\vspace{-0.2cm}
\caption{\small 
$\eps$-Convergence is reached when the agents fall within groups 
with disjoint enclosing intervals of length
at most $\eps$ and no further interaction ever 
takes place between distinct groups.
\label{fig-convergedef}}
\end{figure}

\subsection{Convergence}\label{sec-convergence}

Bounding the convergence rate of agreement systems faces the obvious
difficulty that an adversary can always make the
temporal network free of any edges joining distinct nodes
for as long as it wants, and then, at some point far into the 
future, add all $\binom{n}{2}$ edges permanently to the temporal network
in order to make all the agents cluster around the same point. 
How then can one hope to bound the convergence time since it can be
arbitrarily large yet finite? The total $s$-energy 
is meant to deal with precisely that problem.
Given $0<\eps<1/2$, we say that
a step $t$ is {\em trivial} (where $\eps$ is understood)
if all the edges of $G_t$ have length at most $\eps$.
Recall that $G_t$ always has $n$ self-loops.\footnote{A self-loop
always has zero length but an edge $(i,j)$ may be of
zero length without being a self-loop.}
The {\em communication count}
$C_\eps$ is defined as the total number of nontrivial steps.
Intuitively, it is a way of ignoring microscopic motions.
The system is said to {\em $\eps$-converge}
if the $n$ agents can eventually be partitioned into subsets 
with disjoint enclosing intervals of length
at most $\eps$ and, from that point on, no further interaction ever 
takes place between distinct subsets (Figure~\ref{fig-convergedef}).
{\em Consensus} refers to the case of a one-set partition.
Visually, $\eps$-convergence means that the system 
eventually freezes. We have this obvious relation between $\eps$-convergence
and triviality:

\begin{fact}\label{eps-convergence}
$\!\!\! .\,\,$
An $n$-agent bidirectional agreement system 
$(n-1)\eps$-converges by the time its last nontrivial step
has elapsed.
If the temporal network remains connected at
all times, then the system $(n-1)\eps$-converges
to consensus within $C_{\eps}$ time.
\end{fact}

\smallskip

\begin{theorem}\label{stoptime}
$\!\!\! .\,\,$
The maximum communication count $C_\eps(n)$ of
any $n$-agent bidirectional agreement system
with unit initial diameter satisfies
$$
O(\rho)^{-\lfloor n/2\rfloor}
\log \hbox{$\frac{1}{\eps}$} 
\leq C_\eps (n) \leq 
\min\Bigr\{\,
\hbox{$\frac{1}{\eps}$}\,
 \rho^{-O(n)}\, , \,
(\log \hbox{$\frac{1}{\eps}$})^{n-1}
\rho^{-n^2-O(1)}\, \Bigr\}.
$$
\end{theorem}

\medskip
\noindent
If the initial diameter $D$ is not 1, then
we must replace $\eps$ by $\eps/D$ in
the bounds for $C_\eps(n)$.
We easily check that the bound is essentially tight as
long as $\eps$ is not superexponentially small. Indeed,
for any constant $a>0$,
there exist two constants $b,c>0$ such that,
if $\eps \geq \rho^{an}$, then (i) the communication
count is at most $(1/\rho)^{bn}$ and (ii) there exists an 
agreement system and a starting configuration
for which the communication count
exceeds $(1/\rho)^{cn}$. Put more succinctly,

\smallskip

\begin{corollary}\label{stoptime-coro}
$\!\!\! .\,\,$
If $\eps\geq \rho^{O(n)}$, then $C_\eps(n)= \rho^{-\Theta(n)}$.
\end{corollary}

\begin{theorem}\label{stoptime-reversible}
$\!\!\! .\,\,$
For any $0<\eps<\rho/n$, an $n$-agent reversible agreement system
$\eps$-converges to consensus in time 
$O(\frac{1}{\rho}\, n^2\log \frac{1}{\eps})$, 
\end{theorem}

\smallskip\smallskip

Let $\delta= \max_{i,t} |N_i(t)|$ be the 
maximum degree of any node in the temporal network.
The assignment $q_i=\delta$ is valid and gives 
$\rho= 1/\delta$.  The theorem implies $\eps$-convergence
to consensus in $O(\delta n^2\log \frac{1}{\eps})$ time.
A similar result holds if the degree of any given node
does not change over time and each agent moves
to the mass center of its neighbors: $x_i(t+1)= 
(1/|N_i(t)|) \sum_{j\in N_i(t)} x_j(t)$.
If we now consider the case of a time-invariant graph,
we retrieve the usual polynomial mixing time
bound for lazy random walks in undirected graphs.

The communication count is related to the total $s$-energy
via the obvious inequality: 
$$C_\eps\leq \eps^{-s} E(s).$$
In view of this relation,
the two upper bounds in Theorem~\ref{stoptime}
follow directly from those in Theorem~\ref{En(s)bound}:
simply set $s= 1$ and $s=n/\ln\frac{1}{\eps}$,
respectively.
Note that the second assignment can be assumed to
satisfy $s<1$, since it only concerns the case where
$\frac{1}{\eps}\rho^{-O(n)}$ is the bigger term
in the right-hand side of the expression in Theorem~\ref{stoptime}.
For reversible systems, we set $s=1/\ln\frac{1}{\eps}$,
and observe that the number of steps witnessing
a diameter in excess of $\eps$ is at most
$\eps^{-s}E^D_n(s)= O(\frac{1}{\rho}\, n^2\log \frac{1}{\eps})$.
This bounds the time it takes for the diameter to dip below $\eps$
and stay there forever (since it cannot grow); 
hence Theorem~\ref{stoptime-reversible}.
\hfill $\Box$
\proofend

\section{Applications}

We highlight the utility of the total $s$-energy
by looking at five examples:
opinion dynamics~(\S\ref{opinion-dynamics});
social epistemology~(\S\ref{truthseeksys});
Kuramoto synchronization~(\S\ref{sync});
bird flocking~(\S\ref{flocking});
and products of stochastic matrices~(\S\ref{prodstoch}).

\subsection{Opinion Dynamics}\label{opinion-dynamics}

The Krause opinion dynamics model~\cite{hegselmanK, krause00}
is a sociological framework for tracking opinion polarization
in a population. In its $d$-dimensional version,
the {\em bounded-confidence model}, as it is often called,
sets a parameter $0<r<1$ and, at time $0$,
specifies the opinions of $n$ agents
as $n$ points in the unit cube $[0,1]^d$.
At time $t\geq 0$, each opinion $x$ moves to the
position given by averaging all the opinions
that happen to fall within
the Euclidean ball centered at $x$ of radius $r$
(or some other shape).
Viewed as a multiagent agreement system,
$G_t$ consists of $n$ nodes (the agents) with edges
joining any two of them within distance $r$ of each other.
The dynamics is specified by
\begin{equation}\label{opinion-dyn}
x_i(t+1)= \frac{1}{|N_i(t)|}\sum_{j\in N_i(t)} x_j(t),
\end{equation}
where $N_i(t)$ is the set of neighbors of node $i$ in $G_t$,
which as usual includes $i$ itself.
The system is known to 
converge~\cite{blondelHT09,krause00,lorenz03,lorenz05}.
Theorem~\ref{stoptime} allows us 
to bound how long it takes to reach equilibrium.
Consider a Cartesian coordinate system.
In view of~(\ref{pij-rho}, \ref{opinion-dyn}), 
we may set $p_{ij}(t)= 1/|N_i(t)|$ and $\rho=1/n$ to
make the opinion dynamics system along each coordinate axis 
conform to a one-dimensional multiagent agreement model~(\ref{gen-dyn}).
We can assume that the maximum diameter $D$ along each axis 
is at most $r n$ at time $0$ and hence thereafter.
Indeed, by convexity, if along any coordinate axis
the $n$ opinions have diameter
greater than $r n$, then they can be split into
two subsets with no mutual interaction now and forever.
Set $\eps = r/2$ and let $t_\eps$ be
the smallest $t$ such that $G_t$ consists only
of edges in ${\mathbb R}^d$ of length at most $\eps$. During
the first $dC_{\eps/\sqrt{d}\,}(n)+1$ steps, it must be
the case that, at some time $t$, the graph $G_t$ contains
only edges of length at most~$\eps$.
By Theorem~\ref{stoptime}, therefore

\begin{equation}\label{teps-opinion}
t_\eps\leq d^{3/2} \frac{D}{\eps}\, n^{O(n)}= 
2d^{3/2} \,n^{1+O(n)}= n^{O(n)}.
\end{equation}
Each connected component of $G_{t_\eps}$ is a complete
graph. To see why, observe that if opinion
$x$ is adjacent to $y$ in $G_{t_\eps}$ and the same is
true of $y$ and $z$, then $x$ and $z$ are at a distance
at most $2\eps= r$, hence are connected
and therefore at distance at most $\eps$ at time $t_\eps$.
This ``transitive closure'' argument
proves our claim. This implies that 
the opinions within any connected component 
end up at the same position at time $t_\eps +1$.
Of course, when two opinions are joined together
they can never get separated. 
The argument is now easy to complete.
Either $G_{t_\eps}$ consists entirely of isolated nodes,
in which case the system is frozen in place, or it consists
of complete subgraphs that collapse into single points.
The number of distinct opinions decreases by at least one,
so this process can be repeated at most $n-2$ times.
By~(\ref{teps-opinion}), this proves that 
Krause opinion dynamics converges in $n^{O(n)}$ time.
We summarize our result.

\smallskip

\begin{theorem}\label{opinion-dyn-thm}
$\!\!\! .\,\,$
Any initial configuration of $n$ opinions in the bounded-confidence Krause model
with equal-weight averaging converges to a fixed configuration
in $n^{O(n)}$ time. 
\end{theorem}

Martinez et al~\cite{martinezBCF07}
have established a polynomial bound for
the one-dimensional case, $d=1$. 
While extending their proof technique to higher dimension
might be difficult, 
a polynomial bound could well hold
for any constant $d$. We leave this as an
interesting open problem.


\subsection{Truth-Seeking Systems}\label{truthseeksys}

In their pioneering work in
computer-aided social epistemology,
Hegselmann and Krause considered a variant
of the bounded-confidence model that assumes 
a {\em cognitive division of labor}~\cite{hegselmanK2006}.
The idea is to take the previous model and fix one agent, {\em the truth}, 
while keeping the $n-1$ others mobile.
A {\em truth seeker} is a mobile agent 
joined to the truth in every $G_t$.
All the other mobile agents are {\em ignorant},
meaning that they never connect to the truth via an edge,
although they might indirectly communicate with it
via a path.
Any two mobile agents are joined in $G_t$ whenever
their distance is less than $r$. (Using open balls
simplifies the proofs a little.)
Hegselmann and Krause~\cite{hegselmanK2006} 
showed that, if all the mobile agents
are truth seekers, they eventually reach
consensus with the truth. 
Kurz and Rambau~\cite{kurzR} proved that
the presence of ignorant agents 
cannot prevent the truth seekers from
converging toward the truth. The proof is quite technical
and the authors leave open the higher-dimensional case.
We generalize their results to any dimension and, as a bonus,
bound the convergence rate. 
\smallskip

\begin{theorem}\label{truth-seeking-thm}
$\!\!\! .\,\,$
Any initial configuration of $n$ opinions in ${\mathbb R}^d$ in the 
truth-seeking model converges, with all
the truth seekers coalescing around the truth.
If, in addition, we assume that the initial coordinates of each opinion 
as well as the radius $r$
are encoded as $O(n)$-bit rationals then, 
after $n^{O(n)}$ time,
all the truth seekers lie within a ball
of radius $2^{-n^{cn}}$ centered at the truth, for any 
arbitrarily large constant $c>0$. Ignorant agents either lie
in that ball or are frozen in place forever.
This holds in any fixed dimension.
\end{theorem}

\proof
Along each coordinate axis, 
a truth-seeking system falls within the fixed-agent
agreement model and, as we saw in~\S\ref{multiagent-dynamics},
can be simulated by a $(2n-1)$-agent one-dimensional bidirectional
agreement system with at most twice the initial diameter. (Note that
the $2n-1$ agents do not form a truth-seeking system because there
are no edges connecting the group of $n$ original agents to its reflection.)
Convergence follows from Fact~\ref{eps-convergence}.
As we observed in the previous section,
restricting ourselves to the equal-weight
bounded confidence model allows us to set 
$\rho=1/(2n-1)$.
(We could easily handle more general
weights but this complicates the notation
without adding anything of substance to the argument.)
Kurz and Rambau~\cite{kurzR} observed that the convergence
rate cannot be bounded as a function of $n$ and $\rho$ alone
because it also depends on the initial conditions (hence
the need to bound the encoding length of the initial coordinates).

Set $\eps=2^{-bn}$ for some large enough constant $b>0$, and
define $t_\eps$ as
the smallest $t$ such that $G_t$ consists only of
edges not longer than $\eps$. 
By the same projection argument we used in~(\ref{teps-opinion})
and the observation that the initial diameter is $2^{O(n)}$,
$t_\eps= n^{O(n)}$.
The subgraph of $G_{t_\eps}$ induced by the mobile agents
consists of disjoint complete subgraphs. Indeed,
the transitive closure argument of the previous section
shows that the distance between any two agents within the
same connected component is at most
$2\eps= 2^{1-bn} < r$ (the inequality
following from the $O(n)$-bit encoding of $r$),
hence at most $\eps$.
For similar reasons, 
the truth agent cannot join more than one of these complete subgraphs
(referring here and below to the original system and {\em not} the duplicated version);
therefore, all the subgraphs consist of ignorant agents, except for one
of them, which contains all the truth seekers and to which 
the truth agent is joined. This {\em truth group} might contain
some ignorant agents as well, ie, mobile agents
not connected to the truth. For that reason, the truth group,
in which we include the truth, is a connected subgraph that
might not be complete.
At time $t_\eps +1$, 
the truth group has collapsed into
either a single edge with the truth at one end
or a collinear 3-agent system consisting
of the truth, a truth seeker, and an ignorant agent.
(We refer to a single agent or truth seeker although
it may be a collection of several of them collapsed into one.)
All the other complete subgraphs collapse 
into {\em all-ignorant} single agents.
By Theorem~\ref{opinion-dyn-thm},
there is a time 
\begin{equation}\label{t_0-truth}
t_0=t_\eps+ n^{O(n)}= n^{O(n)}
\end{equation}
by which the all-ignorant agents
will have converged into frozen positions
unless they get to join with
agents in the truth group at some point. 

\bigskip
\noindent
{\sc Case I.}\ 
Assume that the all-ignorant agents do not join with
any agent in the truth group at any time $t> t_\eps$:
The truth group then behaves like a one-dimensional fixed-agent
system with 2 or 3 agents embedded in ${\mathbb R}^d$.
We assume the latter,
the other case being similar, only easier.
We saw in~\S\ref{multiagent-dynamics} how such a system can
be simulated by a one-dimensional $5$-agent bidirectional system of 
at most twice the diameter. 
Recall that agents may represent the collapse of several of them,
so we must keep the setting $\rho=1/(2n-1)$.
The $5$-agent system remains
connected at all times (since its diameter cannot grow); therefore,
by Fact~\ref{eps-convergence} and Theorem~\ref{stoptime},
it $\beta$-converges to consensus by (conservatively) time 
$t_0 + n^{O(1)}(\log\frac{1}{\beta})^4$ time.
By~(\ref{t_0-truth}), this
implies that, for any fixed $c_0>0$,
the agents of the truth group
are within distance $2^{-n^{c_0n}}$ of the truth
after $n^{O(n)}$ time.

\bigskip
\noindent
{\sc Case II.}\
Assume now that an all-ignorant agent $z$
joins with an agent $y$ of the truth group
at time $t_1$ but not earlier in $[t_\eps,t_1)$.
That means that the distance $\|y(t_1)z(t_1)\|_2$
dips below $r$ for the first time
after $t_\eps$. We want to show that $t_1\leq t_0+ n^{O(n)}$,
so we might as well assume that $t_1>t_0$.
Recall that $t_0$ is an upper bound on the time by which
the all-ignorant agents would converge if they never interacted
again with the truth group past $t_\eps$.
Let $L$ be the line along which the truth group evolves
and let $\sigma$ be its (nonempty) intersection with the 
open ball of radius $r$ centered at $z(t_1)=z(t_0)$.
Note that $\sigma$ cannot be reduced to a single
point. This implies that
the shortest {\em nonzero} distance $\Delta$
between the truth and the two endpoints of $\sigma$
is well-defined. (By definition, if the truth sits at 
one endpoint, $\Delta$ is determined by the other one.)
We claim that 
\begin{equation}\label{delta2^n}
\Delta\geq 2^{-n^{O(n)}}.
\end{equation}
Here is why.
It is elementary to express $\Delta$ as a feasible value
of a variable in a system of $m$ linear and quadratic
polynomials over $m$ variables, where $m$ is a
constant (depending on $d$).
The coefficients of the polynomials can be chosen to be
integers over $\ell= n^{O(n)}$ bits. (We postpone
the explanation.) 
We need a standard root separation bound~\cite{yap00}.
Given a system of $m$ integer-coefficient polynomials in $m$ variables
with a finite set of complex solution points, 
any nonzero coordinate has modulus at least 
$2^{-\ell \gamma^{O(m)}}$, where $\gamma-1$ is the maximum degree of any polynomial
and $\ell$ is the number of bits needed
to represent any coefficient. This implies 
our claimed lower bound of $2^{-n^{O(n)}}$ on $\Delta$.

Why is $\ell= n^{O(n)}$? 
At any given time, consider the rationals describing
the positions of the $n$ agents and put them in a form
with one common denominator.
At time $0$, each of the initial
positions now requires $O(n^2)$ bits (instead of just $O(n)$ bits).
A single time step produces
new rationals whose common denominator is
at most $n!$ times the previous one, while
the numerators are sums of at most $n$ previous numerators,
each one multiplied by an integer at most $n!$.
This means that, at time $t$, none of the
numerators and denominators require more than $O(n^2+ tn\log n)$ bits.
The system of equations expressing $\Delta$
can be formulated using integer coefficients
with $O(n^2+ t_0n\log n)$ bits, hence
the bound of $\ell= n^{O(n)}$.
Next, we distinguish between two cases.

\begin{itemize}
\item
{\em The truth is not an endpoint of $\sigma$}:
Then there is a closed segment of $L$ centered at the truth
that lies either entirely outside of $\sigma$
or inside of it. By~(\ref{delta2^n}), 
the segment can be chosen of length at least $2^{-n^{O(n)}}$.
Setting $c_0$ large enough, as we saw earlier,
the agents of the truth group
are within distance $2^{-n^{c_0n}}$ of the truth
after $n^{O(n)}$ time; therefore,
$t_1\leq t_0 + n^{O(n)}$, else the diameter of the truth
group becomes too small to accommodate $\Delta$.
\item
{\em The truth is an endpoint of $\sigma$}:
Quite clearly, 
$\beta$-convergence alone does not suffice to
bound $t_1$; so we reason as follows.
When the truth
group has $\beta$-converged (for the previous value
of $\beta$), the only way its mobile
agents avoided falling within $\sigma$ (in which case
the previous bound on $t_1$ would hold) is if
the truth group ended up separated from $\sigma$ by the truth
(lest one of the mobile agents lay in $\sigma$).
By convexity, however, this property remains true from then on,
and so $z$ can never join $y$, which contradicts our assumption.
\end{itemize}

\noindent
When agents $y$ and $z$ join in $G_t$ at time $t=t_1$,
their common edge is of length at least $r/3$ 
unless $y$ or $z$ has traveled a distance
at least $r/3$ between $t_\eps$ and $t_1$.
In all cases, the system must expend 1-energy at least
$r/3$ during that time interval. By Theorem~\ref{En(s)bound},
this can happen at most $n^{O(n)}(3/r)= n^{O(n)}$ times.
We can repeat the previous argument safely each time,
even though the bit lengths will increase.
At the completion of this process, we are back to Case I. 
\hfill $\Box$
\proofend

\vspace{0cm}
\begin{figure}[htb]
\begin{center}
\hspace{.2cm}
\includegraphics[width=6.5cm]{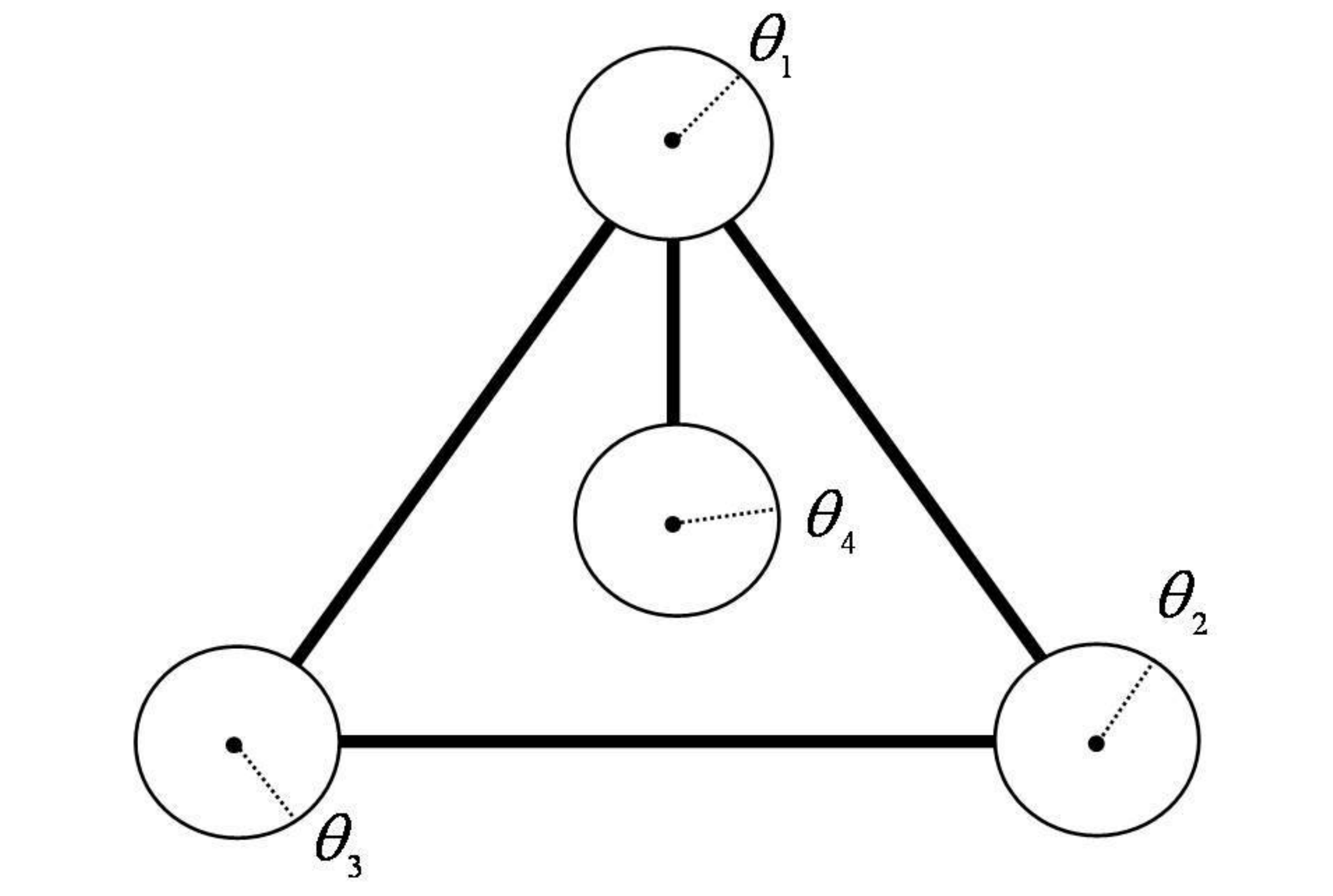}
\end{center}
\caption{\small Four coupled oscillators connected by four edges.
\label{fig-kuramoto}}
\end{figure}

\subsection{Kuramoto Synchronization}\label{sync}

The {\em Kuramoto model} is a general framework for nonlinear
coupled oscillators, with a dazzling array of 
applications: circadian neurons, chirping
crickets, microwave oscillators, yeast cell
suspensions, pacemaker cells in the heart, etc.
Winfree's pioneering work on the subject led
Kuramoto to formulate the standard sync model
for coupled oscillators~\cite{strogatz00,winfree67}.
The system consists of $n$ oscillators:
the $i$-th one has phase $\theta_i$ and
natural frequency $\omega_i$.
In its original formulation, the model is 
a mean-field approximation that assumes all-pair coupling.
A more realistic assumption is to use a time-varying
network to model communications. 
Considerable work has been done on this problem;
see~\cite{earlS,jadbabaieMB,
liXLY,martinez09,mcgrawM,papacJ05,papacJ06,wangSCF,zanette04} for a small sample.
Further research introduced a time-1 discretization
of the continuous model~\cite{martinez09,Moreau2005,scardoviSS,triplettKM}.
Assuming all oscillators share the same natural frequency,
a fixed phase shift yields the dynamics:
\begin{equation*}
\theta_i(t+1)= \theta_i(t) 
+ \frac{K\Delta T}{|N_i(t)|}
\sum_{j\in N_i(t)}\sin(\theta_j(t)-\theta_i(t)),
\end{equation*}
where $|N_i(t)|$ is the degree of $i$ in 
the communication graph $G_t$, which, as always,
counts the self-loop at $i$ (Figure~\ref{fig-kuramoto}).
As in~\cite{Moreau2005}, we also assume
that all the agents' phases start in
the same open half-circle. By shifting the origin,
we express this condition as
$\alpha-\pi/2\leq \theta_i(0)\leq \pi/2$, for
some arbitrarily small positive constant $\alpha$.
This implies that 
$$\sin(\theta_j(0)-\theta_i(0)) = a_{ij}(\theta_j(0)-\theta_i(0)),$$
for $\alpha/4\leq a_{ij}\leq 1$.
By~(\ref{pij-rho});
therefore, to make the dynamics conform
to a bidirectional multiagent agreement system at time $0$,
it suffices to enforce the constraints:
$$
\frac{4n\rho}{\alpha} \leq K\Delta T \leq 1-\rho\, .
$$
Choosing $\rho=b\alpha/n$ for a small enough constant $b>0$,
we note that the constraints are roughly
equivalent to $0< K\Delta T<1$. 
By convexity, the angles at time $1$
remain within $[\alpha-\pi/2, \pi/2]$;
therefore, our previous argument
can be repeated to show that the synchronization dynamics
fits within the bidirectional agreement model at all times.
The result below follows from Corollary~\ref{stoptime-coro}.
We note that it is impossible to bound the actual time to convergence
unless we make assumptions about the temporal network.

\smallskip

\begin{theorem}\label{kuramoto-thm}
$\!\!\! .\,\,$
Any Kuramoto synchronization system
with $n$ oscillators sharing the same natural frequency
and initialized in an open half-circle
$\eps$-converges after $n^{O(n)}$ nontrivial steps,
for any $\eps> n^{-cn}$ and any constant $c>0$.
This holds regardless of the temporal communication network.
\end{theorem}

\smallskip

\subsection{Bird Flocking}\label{flocking}

Beginning with Reynolds's pioneering work
in the mid-eighties, 
bird flocking has generated an abundant literature, with a sudden
flurry of interest 
in the last few years. 
Mathematically, flocking appears more complex than
the previous agreement systems because the averaging
and the communications do not operate over precisely
the same objects: it is the velocities that are averaged but
the positions (ie, the integrals of the velocities)
that determine the temporal network.
Many models have been studied in the literature but most
of them are variants of the 
following~\cite{chazFlockSODA, CuckerSmale1, jadbabaieLM03, vicsekCBCS95}:
given the initial conditions $z(0)$ and $z(1)$, for any $t>0$,

\begin{equation}\label{modelD}
\begin{cases}
\, z(t)= z(t-1)+ v(t); \\
\, v(t+1)= P(t) v(t).
\end{cases}
\end{equation}
The vectors $z(t), v(t)$ encode the positions
and velocities of the $n$ birds in ${\mathbb R}^3$:
each coordinate of $z(t)$ and $v(t)$ is itself a
three-dimensional vector. (These vectors
are often expressed in ${\mathbb R}^{3n}$ via a tensor product;
the notation here is easier as long as one remembers that
the coordinates are themselves three-dimensional vectors.)
The $n$-by-$n$ stochastic matrix $P(t)$ has nonzero
diagonal entries and its other positive entries
correspond to the edges of $G_t$; the communication
graph $G_t$ links any two birds within a fixed distance of 
each other.
Intuitively, each bird averages out its own velocity with
those of its neighbors in $G_t$: all of its neighbors weigh equally
in the average except perhaps for itself, ie, for
fixed $i$, all nonzero
$p_{ij}(t)$'s are equal, with the possible
exception of $p_{ii}(t)$; all the entries in
$P(t)$ are rationals over $O(\log n)$ bits.

It suffices to set $\rho=n^{-b}$, for 
a large enough constant $b>0$, to make flocking conform to 
the bidirectional multiagent agreement model,
with $v(t)$ encoding into a single vector
the $n$ points $(x_1(t),\ldots, x_n(t))$.
By Corollary~\ref{stoptime-coro},
the system $\eps$-converges within $n^{O(n)}$
nontrivial steps for $\eps\geq n^{-cn}$
and any constant $c>0$.
We showed in~\cite{chazFlockPaper} that the  
sequence $G_t$ always converges to a fixed graph $G$,
but that the number of steps to get there can be
astronomical: it can be as high
as a tower-of-twos of height on the order of $\log n$, which,
amazingly, is tight.

\smallskip

\begin{theorem}\label{flocking-thm}
$\!\!\! .\,\,$
The velocities of $n$ birds $\eps$-converge 
after $n^{O(n)}$ nontrivial steps,
for any $\eps> n^{-cn}$ and any constant $c>0$.
The number of steps prior to the convergence
of the temporal network to a fixed graph
is no higher than a tower-of-twos
of height $O(\log n)$; this bound is optimal in the worst case.
\end{theorem}

\subsection{Products of Stochastic Matrices}\label{prodstoch}

Let ${\mathcal P}$ be the family of
$n$-by-$n$ stochastic matrices such that
each $P\in {\mathcal P}$ satisfies the three standard constraints:
(i) {\em self-confidence} (nonzero diagonal entries);
(ii) {\em mutual confidence} (no pair $p_{ij}, p_{ji}$
with exactly one 0); 
and (iii) {\em nonvanishing confidence} 
(positive entries at least $\rho$).  
Lorenz~\cite{lorenz05}
and Hendrickx and Blondel~\cite{HendrickxB}
independently proved the following counterintuitive result:
in any finite product of matrices in ${\mathcal P}$,
each nonzero entry is at least $\rho^{O(n^2)}$.
What is surprising is that this lower bound is
uniform, in that it is independent
of the number of multiplicands in the product.
We improve this lower bound to 
its optimal value:

\smallskip

\begin{theorem}\label{matrixlb-thm}
$\!\!\! .\,\,$
Let ${\mathcal P}$ be the family of
$n$-by-$n$ real stochastic matrices such that
any $P\in {\mathcal P}$ satisfies: 
each diagonal entry is nonzero; 
no pair $p_{ij}, p_{ji}$ contains exactly one zero;
and each positive entry is at least $\rho$.
In any finite product of matrices in ${\mathcal P}$,
each nonzero entry is at least $\rho^{n-1}$.
The bound is optimal.
\end{theorem}

\section{The Proofs}\label{Proofs}

It remains for us to prove:
Theorem~\ref{En(s)bound} (upper bound
for $s=1$ in~\S\ref{generalcase-s=1}, upper bound for
$s<1$ in~\S\ref{generalcase-s<1}, and lower bounds
in~\S\ref{lowerbounds}),
Theorem~\ref{matrixlb-thm}~(\S\ref{generalcase-s<1}),
Theorem~\ref{En(s)reversible}~(\S\ref{reversiblecase}),
and the lower bound of Theorem~\ref{stoptime}~(\S\ref{lowerbounds}).

\subsection{The General Case: $s=1$}\label{generalcase-s=1} 

We prove the upper bound of Theorem~\ref{En(s)bound} for $s=1$.
We show that $E_n(1)\leq \rho^{-O(n)}$ by bounding the
kinetic $s$-energy.

\smallskip
\paragraph{Wingshift systems.}

We introduce a {\em wingshift system}, which provides
a simpler framework for the proof.
Since we focus on a single transition at a time,
we write $a_i,b_i$ instead of $x_i(t),x_i(t+1)$
for notational convenience, and we
relabel the agents so that $0\leq a_1\leq\cdots\leq a_n\leq 1$.
Given $a_1,\ldots, a_n$, the agents move to their
next positions $b_1,\ldots, b_n$ 
and then repeat this process endlessly 
in the manner described below. 
Let $\ell(i)$ and $r(i)$ be indices satisfying the following inequalities:

\begin{itemize}
\item[]\hspace{1cm}{\sc rule} 1: \ \
$1\leq \ell(i)\leq i\leq r(i)\leq n$
and 
$(\ell \circ r)(i)\leq i\leq (r \circ \ell)(i)$;
\item[]\hspace{1cm}{\sc rule} 2: \ \ 
$a_{\ell(i)}+\delta_i\leq b_i\leq a_{r(i)}-\delta_i$,
where $\delta_i= \rho (a_{r(i)}- a_{\ell(i)})$.
\end{itemize}
Each agent $i$ picks an {\em associate} to its left
(perhaps itself)
and one to its right, $\ell(i)$ and $r(i)$, respectively.
It then shifts anywhere in the
interval $[a_{\ell(i)}, a_{r(i)}]$, 
though keeping away from the endpoints
by a small distance $\delta_i$. This process
is repeated forever, with each agent given a chance
to change associates at every step.
Any multiagent
agreement system with parameter $\rho$ can be modeled
as a wingshift system: each agent picks its leftmost and
rightmost neighbors as associates; note that the wingshift graph is
sparser but now dependent on the embedding.
Bidirectionality ensures rule 1: it says that
the interval $[\ell(i),r(i)]$ should contain $i$ as well as
all agents $j$ pointing to $i$.\footnote{This is necessary
for convergence. 
Consider three agents 
$a_1=0$, $a_2=\frac{1}{2}$, $a_3=1$, with
$\ell(1)=r(1)=\ell(2)=1$ and $r(2)=\ell(3)=r(3)=3$.
Agents $1$ and $3$ are stuck in place while
agent $2$ can move about freely forever.}
By analogy, we define the total $1$-energy 
of the wingshift system as
$V= \sum_{t\geq 0} V_t$, where (with $a_i$ denoting $x_i(t)$)
$V_t= \sum_{i=1}^n \, (a_{r(i)}- a_{\ell(i)})$. 
The desired upper bound $E_n(1)= \rho^{-O(n)}$ follows
trivially from this bound on $V$:

\begin{theorem}\label{V-wingshiftUB}
$\!\!\! .\,\,$
The maximal total $1$-energy of an $n$-agent 
wingshift system with unit initial diameter
and parameter $\rho$ is at most $\rho^{-O(n)}$.
\end{theorem}

\vspace{0.5cm}
\begin{figure}[htb]
\begin{center}
\hspace{.2cm}
\includegraphics[width=10cm]{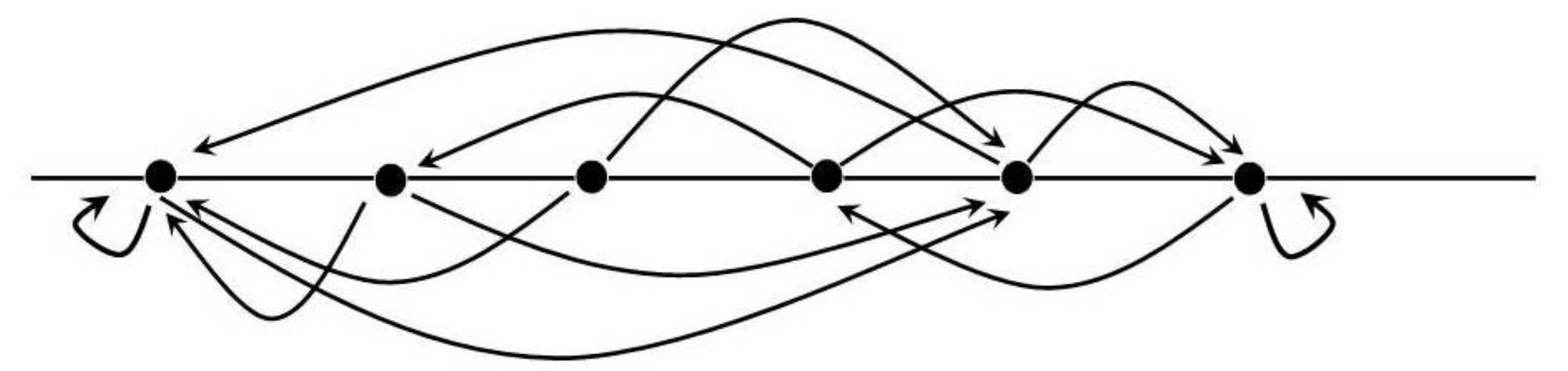}
\end{center}
\vspace{-0.2cm}
\caption{\small A six-node wingshift system.
\label{fig-wingshift}}
\end{figure}
\vspace{0.5cm}

\smallskip

As usual, we assume that $\rho$ is smaller than a suitable constant.
We need some notation to describe rightward paths
in the wingshift system: 
$r(i,0)=i$ and $r(i,k)=r(r(i,k-1))$ for $k>0$.
We define the distance between an agent
and its right associate, $\Delta_i= a_{r(i)}-a_i$.
When traversing a rightward path 
$i=r(i,0),r(i,1),\ldots, r(i,k)$ etc, a sudden drop
in $\Delta_{r(i,k)}$ is of particular interest; so we introduce
$r_i= r(i,k_i)$, where
$$
k_i=\min\,\Bigl\{\, k\geq 0\, |\, \Delta_{r(i,k)}
  \geq \hbox{$\frac{2}{\rho}$} \,\Delta_{r(i,k+1)}\,\Bigr\} \, .$$
The agent $r(r_i)$ is called the {\em stopper} of $i$. Our interest
in stoppers is that the nontrivial ones always move left, thus pointing
to some obligatory motion in the system at the time step under consideration.

\vspace{0.5cm}
\begin{figure}[htb]
\begin{center}
\hspace{.2cm}
\includegraphics[width=7cm]{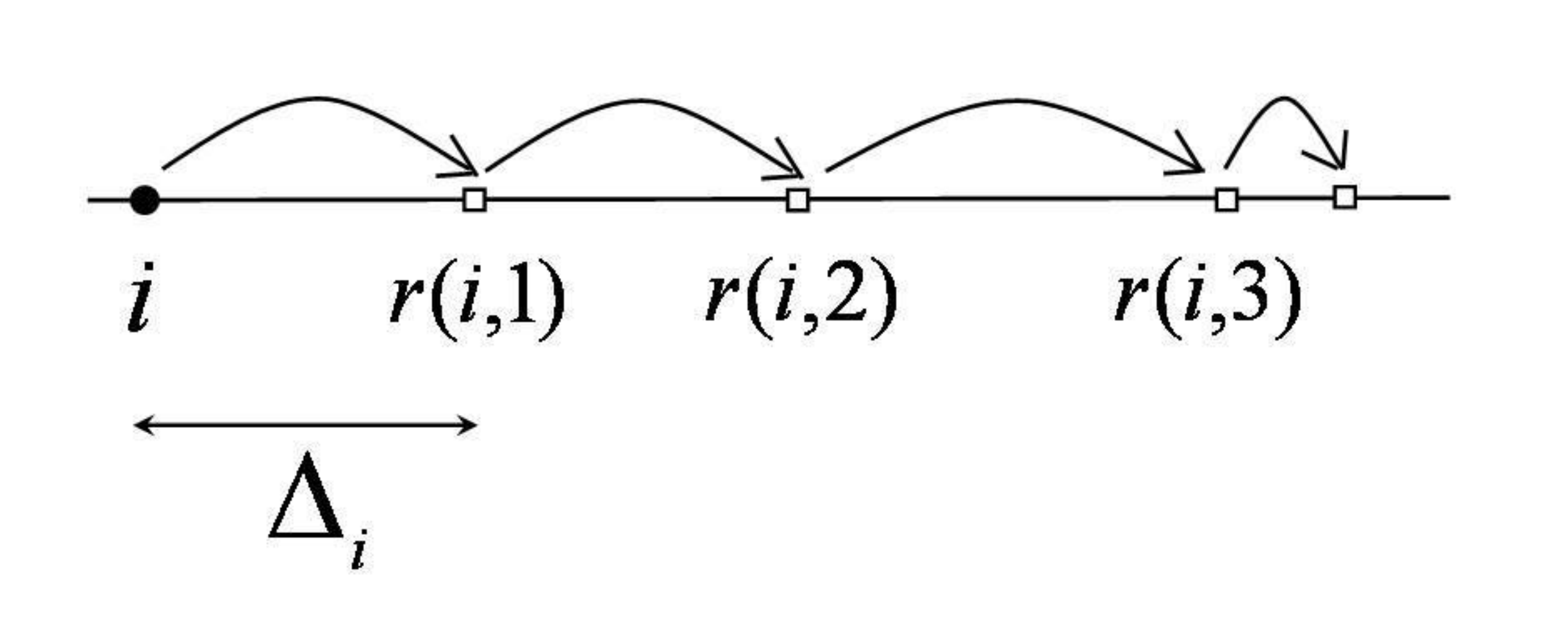}
\end{center}
\vspace{-0.8cm}
\caption{\small The stopper of agent $i$ is $r(i,3)$; $r_i= r(i,2)$; and $k_i=2$.
\label{fig-stopper}}
\end{figure}
\vspace{0.5cm}

\begin{lemma}\label{abDelta}
$\!\!\! .\,\,$
The stopper $u$ of any agent $i$ 
satisfies
$a_u-b_u \geq (\frac{\rho}{2})^{k_i+1} \Delta_i$.
\end{lemma}
 
\proof
By rule 1,  
$a_{r(u)} - a_{\ell(u)} \geq 
\Delta_{r_i} + \Delta_{u} $.
Since $\Delta_{u} \leq \frac{\rho}{2} \,\Delta_{r_i}$,
it follows by rule 2 that
\begin{equation*}
\begin{split}
b_{u}
&\leq a_{r(u)}- \rho( a_{r(u)} 
- a_{\ell(u)} )
\leq a_{u}+\Delta_{u}
  - \rho( \Delta_{r_i} + \Delta_{u} ) \\
&\leq a_{u}+ ((1-\rho)\rho/2 -\rho) \Delta_{r_i}
\leq  a_{u}- (1+\rho)(\rho/2)^{k_i+1}\Delta_i \, .
\end{split}
\end{equation*}
The last inequality follows from the fact that
$\Delta_{r_i}\geq (\rho/2)^{k_i}\Delta_i$;
it is not strict because both sides are equal if $k_i=0$.
\hfill $\Box$
\proofend

\smallskip
\noindent
We bound $V$ by tallying the kinetic $1$-energy 
of the system, 
$K= \sum_{t\geq 0} K_t$, where 
$K_t= \sum_{i=1}^n |a_i-b_i|$.
Recall that $a_i$ and $b_i$ are the positions
of agent $i$ at times $t$ and $t+1$, respectively.

\begin{lemma}\label{work-motion}
$\!\!\! .\,\,$
If $K$ is finite, then $V\leq  n 2^n \rho^{1-n} K$.
\end{lemma}

\proof
Obviously,
$$
V = \sum_{t\geq 0} \, \sum_{i=1}^n \, (a_{r(i)}- a_{\ell(i)})
\leq \sum_{t,i} 
     \Bigl\{ (a_i-a_{\ell(i)}) + (a_{r(i)}-a_i) \Bigr\}
\leq  \sum_{t,i} \Delta_i + \sum_{t,i} \Delta_i'
 \, , 
$$
where $\Delta_i'= a_i-a_{\ell(i)}$.
By Lemma~\ref{abDelta},
$$
\sum_{t\geq 0} \, \sum_{i=1}^n 
\Delta_i \leq  \sum_{t\geq 0} \, \sum_{i=1}^n \,
 \Bigl(\frac{2}{\rho}\Bigr)^{k_i+1} (a_{r(r_i)}-b_{r(r_i)})
\leq  n \Bigl(\frac{2}{\rho}\Bigr)^{n-1}
         \sum_{t\geq 0} \, \sum_{i=1}^n \, |a_i-b_i| \, .
$$
A mirror-image argument yields the same 
upper bound on $\sum_{t,i} \Delta_i'$.
\hfill $\Box$
\proofend

\smallskip

\paragraph{The idea behind the proof.}

By symmetry, we can assume that at least
half of the contribution to $K$
is provided by rightward motions, ie,
$\frac{1}{2} K\leq 
   \sum_{t,i}\{\, b_i-a_i \,|\, b_i>a_i\,\}$.
Thus we can conveniently ignore all leftward travel
for accounting purposes.
We use an ``amortization'' technique that involves
assigning a {\em credit account} to each agent.
Whenever
the agent moves right it is required to 
pay for its travel cost 
by decreasing its account by an amount 
equal to the distance it travels. Credits are injected
into the system only at time $0$; if all travel is paid
for and no account is overdrawn, then clearly the initial
injection is an upper bound on $\frac{1}{2} K$.
The benefit of this approach is that accounts can borrow
from one another, thus creating an ``economy'' of credits.
The proof takes the form of an algorithm 
that drives the trading in a manner
that keeps all accounts solvent; in other words, 
it is an {\em algorithmic proof}~\cite{chazICS}.

Agent $i$ cannot, in a single step, move to the right
by a distance greater than $\Delta_i$. Lemma~\ref{abDelta} suggests
a paying mechanism by which we charge its stopper $u=r(r_i)$
that travel cost; in other words, the leftward travel of $u$
would pay for the rightward travel of the
agents that claim $u$ as a stopper.
If $u$ moves only to the left then its own travel distance is bounded
by 1, and the charging scheme is essentially sound. But what if
$u$ zigzags left and right? The premise of charging $u$ for
the cost of $i$ is that we know how to bound the cost of $u$.
But, if $u$ moves in both directions, we cannot
bound its cost {\em a priori} (whereas we can if it only travels left).
The solution is to look at $u$'s own stopper $u'$
and charge it. This may, in turn, force $u'$ to charge $u''$, etc.
The ``buck passing'' evolves from left to right, so
it must eventually stop. 
This picture suggests that agents should hold more credits the 
further to the right they are: indeed, our credit invariant will
relate an agent's account to its rank.

\smallskip

\paragraph{The algorithmic proof.}
 
At time $t$, the agents are ordered
as $0\leq a_1 < \cdots < a_n\leq 1$. By using
standard perturbation techniques, we can assume strict inequalities
among all agent positions at all times.
We maintain the following credit invariant: at the
beginning of each time step, every
agent $i$ holds $a_i \alpha^{i}$ 
credits in its account, for some fixed parameter $\alpha$,
where again $a_i$ is shorthand for $x_i(t)$.
By way of illustration, consider the trivial case of
two agents, one at $0$ and the other at $1$, meeting
at $1-\rho$ at the next step (we use ties for convenience).
The system holds $\alpha^2$ at time $0$
and $(1-\rho)(\alpha+ \alpha^2)$ at time $1$.
The difference exceeds the travel cost of $1-\rho$ if
$\alpha$ is sufficiently larger than $1/\rho$.

Our algorithmic proof involves setting up a simple data structure,
a linked list, and moving credits around accordingly to specific rules. 
Let $u$ be a stopper such that $b_u<a_u$. Consider 
the lowest-ranked agent $h$ that claims $u$ as its stopper.
We build a doubly-linked list $L_u$ consisting of $u-h+1$ nodes,
each one corresponding to an agent: $h$ is at the head and $u$ at the tail;
scanning $L_u$ takes us through the agents $h,h+1,\ldots, u$.
The nodes scanned after $v$ are called the {\em antecedents} of~$v$.
The {\em rank} $s(v)$ is $h$ plus 
the number of steps it takes to get from $h$ to $v$ in $L_u$.
Ranks are implied by the list, so that
inserting a node automatically adds one
to the ranks of its antecedents. 
Initially, the rank of the node $v$ corresponding to agent $k$ 
is just $k$ and its {\em position}, denoted by $a(v)$, 
is $a_k$. The node following (resp. preceding) $v$ in $L_u$, if it exists,
is denoted by ${\tt next}(v)$ (resp. ${\tt prev}(v)$).
We identify the node $m$ with the highest-ranked agent such that $a_m<b_u$.
Let $\beta= \rho/2$ and $\alpha= 6/\rho^2$.

\vspace{0.5cm}
\begin{figure}[htb]
\begin{center}
\hspace{-0.2cm}
\includegraphics[width=7cm]{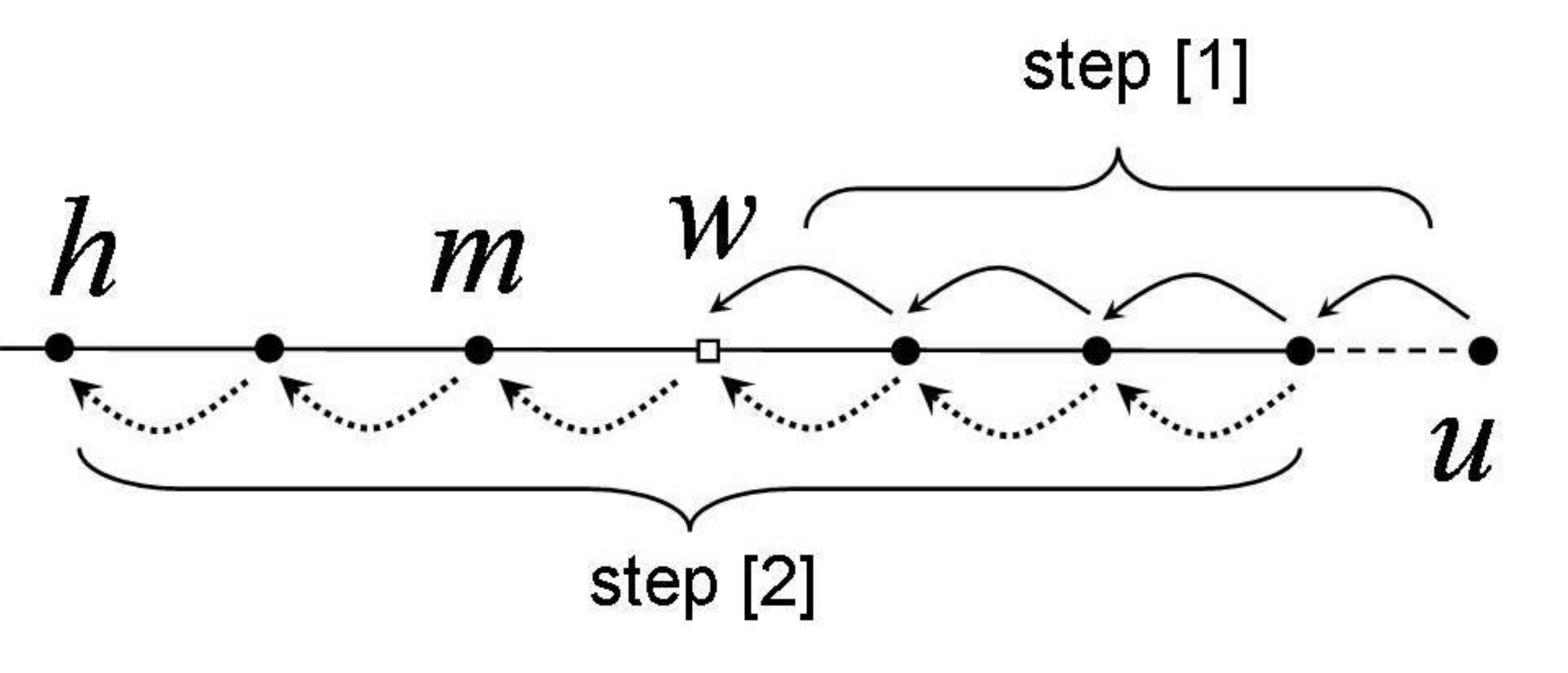}
\end{center}
\vspace{-0.0cm}
\caption{\small Credits are transferred via plain arrows in step [1] 
and dashed arrows in step [2]. Position $a(v)= a_h, a_m, b_u, a_u$ for
$v= h, m, w, u$.
\label{fig-s1proof}}
\end{figure}
\vspace{0.5cm}

\medskip

\vspace{0.3cm}

{\small
\par\medskip
\renewcommand{\sboxsep}{0.7cm}
\renewcommand{\sdim}{0.8\fboxsep}
\hspace{1cm}
\shabox{\parbox{12cm}{
$\bullet$\ For each stopper $u$ from right to left, if $b_u<a_u$, do:
\begin{itemize}
\item[\bf{[1]$\,\,$}]
Insert a new node $w$ into $L_u$ right after $m$:
${\tt next}(w) \leftarrow {\tt next}(m)$ and
${\tt next}(m) \leftarrow w$.
Set $a(w)= b_u$.
For each antecedent $v$ of $w$, transfer 
the account of $v$ to ${\tt prev}(v)$. Delete node $u$ from $L_u$. 
\item[\bf{[2]$\,\,$}]
For each node $v$ from the new tail to ${\tt next}(h)$:
\begin{itemize}
\item[$\bullet$]
transfer $\beta(a_u-a(v))\alpha^{s(v)}$ credits from $v$ to ${\tt prev}(v)$;
\item[$\bullet$]
keep $(\beta a(v) + (1-\beta)a_u)\alpha^{s(v)}$ credits in the account of $v$.
\end{itemize}
\item[\bf{[3]$\,\,$}]
Move from $a_i$ to $b_i$ any agent $i$ claiming $u$ as its stopper,
provided that $b_i>a_i$. Enforce all credit invariants.
\end{itemize}
$\bullet$\ Move from $a_i$ to $b_i$ any nonstopper $i$ such that $b_i<a_i$. 
Enforce all credit invariants.
}}
\par
}
\vspace{1cm}

\paragraph{Step [1]}

Since $b_u<a_u$, node $u$ is the stopper of at least one node
strictly to its left, so $|L_u|>1$ and $m$ is well defined.
By assigning $a(w)= b_u$, in effect we move the stopper $u$
to its new position $b_u$, right after agent $m$ in $L_u$.
Shifting accounts one step backwards gets
$w$ to inherit the account of $m+1$ and the new tail to
receive the credits formerly at $u$. If $m=u-1$, step [1] 
ends with the list in the same state as before except for 
$a({\tt tail})$; otherwise,
the $u-m-1$ antecedents of $w$ see their ranks automatically incremented by one
and, among them, the node for any 
agent $k$ acquires the credit account of $k+1$.
(The alternative of keeping the list intact 
and shifting positions $a(v)$ to the right works but
breaks the immutable correspondence between nodes and agents.)
To summarize, at the end of step [1], any node $v$ in $L_u$ ends up with
$a(v) \alpha^{s(v)}$ credits if $v$ comes before $w$ in the list
and $a({\tt next}(v)) \alpha^{s(v)}$ otherwise.\footnote{By abuse of
notation, $a({\tt next(v)})$ denotes $a_u$ if $v$ is the new tail.}

\paragraph{Step [2]}

We prove that all the credit allocations are feasible.
Suppose that $v$ is either $w$ or an antecedent of $w$. 
Agent $v$ has $a({\tt next}(v)) \alpha^{s(v)}$ credits. 
It receives 
$\beta(a_u-a({\tt next}(v)))\alpha^{s({\tt next}(v))}$ 
credits from ${\tt next}(v)$ (which, by our notational convention, is zero
if $v$ is the new tail); it also
gives away $\beta(a_u-a(v))\alpha^{s(v)}$ credits
and keeps $(\beta a(v) + (1-\beta)a_u)\alpha^{s(v)}$ of them,
for a total need of $a_u\alpha^{s(v)}$.
Since $s({\tt next}(v))=s(v)+1$,
the transaction balances out if
\begin{equation*}
a({\tt next}(v)) \alpha^{s(v)}
+ \beta(a_u-a({\tt next}(v)))\alpha^{s(v)+1}
\geq  a_u\alpha^{s(v)},
\end{equation*}
which holds because $\alpha\beta \geq 1$. 
Suppose now that $v$ comes before $w$ in the list.
The only difference is that $v$ now starts out
with an account worth $a(v) \alpha^{s(v)}$ credits.
The balance condition becomes
$$
a(v) \alpha^{s(v)}
+ \beta(a_u-a({\tt next}(v)))\alpha^{s(v)+1}
\geq 
a_u\alpha^{s(v)},
$$
which is equivalent to
\begin{equation}\label{balcond}
a_u-a(v)\leq \alpha\beta ( a_u-a({\tt next}(v)) ).
\end{equation}
To see why~(\ref{balcond}) holds, 
we turn to the wingshift condition---as, at some point, we must.
Among the agents of $L_u$ claiming $u$ as a stopper
and $v$ as an antecedent, let $z$ be the last one.
(Note that $z$ may not be equal to $v$, but because of $h$ it
is sure to exist.) 
Extending the notation in the obvious way,
by Lemma~\ref{abDelta},
$\Delta_z \leq (\frac{2}{\rho})^{k_z+1} (a_u-b_u)$.
If $k_z=0$, then 
$$ 
\frac{a_u-a(v)}{a_u - a({\tt next}(v))}\leq 
\frac{a_u-a_z}{a_u - b_u} =
\frac{\Delta_z}{a_u - b_u}\leq 
\frac{2}{\rho}< \alpha\beta, $$
which proves~(\ref{balcond}).
If $k_z>0$ then, by the maximality of $z$,
$$ a({\tt next}(v))\leq a_{r(z)} \leq a_{r(r(z))} \leq a_u,
$$ 
and
condition~(\ref{balcond}) follows from
$$ 
\frac{a_u-a(v)}{a_u - a({\tt next}(v))}\leq 
\frac{a_u-a_z}{a_u - a_{r(z)}} =
\frac{\Delta_z+ \Delta_{r(z)}+ a_u - a_{r(r(z))}}
{\Delta_{r(z)} + a_u- a_{r(r(z))}}
\leq 
\frac{\Delta_z+ \Delta_{r(z)}}
{\Delta_{r(z)}}
\leq 1+ \frac{2}{\rho}\, .
$$

\smallskip
\paragraph{Step [3]}

Having shown that all the accounts can afford the amounts
specified in the second bullet of step [2],
we now explain how they can pay for all rightward travel.
We use an accounting trick, which is to move agents 
{\em without crossing}: all agents move continuously, one at a time.
Should agent $i$ bump into agent $j$, the latter completes
the former's journey while $i$ stops; 
the process is repeated at each collision.
The main advantage of this scheme is that
agents now keep their ranks at all times (of course they must
also swap identities, thus becoming {\em virtual} agents).  
Unlike before, a rightward move $a_i\rightarrow b_i$ 
will now entail the motion of all the virtual agents in the interval $[a_i,b_i]$
and not only those claiming $u$ as their stopper.
By step [2], each node $v$ is supplied with
$(\beta a(v) + (1-\beta)a_u)\alpha^{s(v)}$ credits, which is at least
$$
a(v)\alpha^{s(v)}
+ (1-\rho)(a_u - a(v))\alpha^{s(v)}
+ a_u - a(v),
$$
because $s(v)\geq 1$. The three-part sum shows explicitly
why virtual agent $v$, whose rank is now fixed, can move right
by a distance of at least $(1-\rho)(a_u - a(v))$ 
while both maintaining its credit invariant 
and paying (comfortably) for the travel cost.
Virtual agent $v$ never needs to move further right than that.
Why is that? The motion might be generated by $v$ itself (if $u$ is its stopper)
or by the ``push'' from an agent on its left.
Either way, at any instant during the continuous motion
of virtual $v$, there is a causing agent $i$ (perhaps $v$ itself)
whose corresponding interval $[a_i,b_i]$ contains
the position of $v$ at that instant. Our claim follows then
from rule 2. Indeed,
$$
b_i\leq a_{r(i)}- \rho (a_{r(i)}- a_{\ell(i)})
\leq \rho a_{\ell(i)} + (1-\rho) a_{r(i)}
\leq \rho a(v) + (1-\rho) a_u
\leq a(v)+ (1-\rho)(a_u - a(v)) .$$

Returning all agents to their nonvirtual status,
we observe that processing stopper $u$ moves to the right
only the agents that claim it as a stopper.
Treating stoppers $u$ in descending order from right to left 
means that none of the agents with $u$ as their stopper
has yet been moved (either to the left as stoppers
or to the right) by the time we handle $u$.
The last step in the boxed algorithm 
can only release credits---think of virtual agents
to see why---and so, maintaining
the corresponding invariants is immediate.
This allows us to bound the kinetic $1$-energy 
by\footnote{A more sophisticated
argument allows us to lower $\alpha$ to $O(1/\rho)$
and thus reduce the constant in the $O(n)$ exponent to~1;
this sort of finetuning is not needed for the 
purposes of this paper.}
$$\frac{1}{2}K\leq \sum_{i=1}^n x_i(0)\alpha^i 
\leq 2 \alpha^n = \rho^{-O(n)}.$$
Theorem~\ref{V-wingshiftUB} follows now
from Lemma~\ref{work-motion}, which
completes the proof of 
the upper bound of Theorem~\ref{En(s)bound} for $s=1$.
\hfill $\Box$
\proofend

\subsection{The General Case: $s<1$}\label{generalcase-s<1}

We prove the upper bound of Theorem~\ref{En(s)bound} for $0<s<1$.
We show that the total $s$-energy
satisfies the recurrence: 
$E_1(s)=0$ and, for $n\geq 2$,
\begin{equation}\label{E_n}
E_n(s)\leq 
2n E_{n-1}(s) +  
(1-(\rho/2)^n)^s E_n(s)
  + n^3.
\end{equation}
We prove~(\ref{E_n}) ``algorithmically'' by describing
a procedure to track the propagation of information
across the temporal network and monitor its effect
on the geometry of the system.
All agents are initially {\em dry}, except for 
agent $1$, which is {\em wet}. Every time a wet agent
communicates with a dry one, the latter becomes wet.
Once wet, an agent always remains so. Through the communication
provided by the temporal network, water propagates
from agent to agent. Bidirectionality ensures that, when
the water ceases to spread to dry nodes, the interval
spanned by the wet agents will be expected to 
have shrunken a little; in other words, communication
acts as a spring that pulls recipients together.

\bigskip

{\small
\par\medskip
\renewcommand{\sboxsep}{0.7cm}
\renewcommand{\sdim}{0.8\fboxsep}
\hspace{1cm}
\shabox{\parbox{12.9cm}{
\begin{itemize}
\item[\bf{[1]$\,\,$}]
Initially, all 
agents are dry except for agent 1. 
Set $W(0)=\{x_1(0)\}$.
\item[\bf{[2]$\,\,$}]
For $t=0,1, \ldots, \infty$:
\begin{itemize}
\item[\bf{[2.1]}]
    Declare wet any agent adjacent to a wet agent in $G_t$.
\item[\bf{[2.2]}]    
    $W^*(t)\leftarrow W(t)\, \cup $
    $\{$ positions at time $t$ of dry agents just
                turned wet $\}$.
\item[\bf{[2.3]}]    
    Move each agent $i$ from $x_i(t)$ to $x_i(t+1)$. 
     [ {\tt If no newly wet agent, then all motion within
    $W(t)=W(t^*)$ occurs in isolation from the $n- |W(t)|$ other agents.} ]
\item[\bf{[2.4]}]    
    $W(t+1)\leftarrow$ 
     $\{$ positions at time $t+1$ of 
              agents corresponding to $W^*(t)$ $\}$.
\end{itemize}
\end{itemize}
}}

\par\bigskip\medskip
}

\medskip
The set $W(t)$ tracks the positions of the wet agents at time $t$.
The auxiliary set $W^*(t)$ includes
the positions at time $t$ of the agents wet at time $t+1$;
it differs from $W(t+1)$ only in that the latter gives the 
positions at time $t+1$.
Let $\|W(t)\|$ denote the length of the
smallest interval enclosing $W(t)$ and 
let $\{t_k\}_{k\geq 1}$ be the times $t\geq 0$, 
in chronological order, at which $|W^*(t)|> |W(t)|$ 
(ie, at least one dry agent turns wet at time $t$).\footnote{Both
$W(t)$ and $W^*(t)$ are understood as multisets. Note 
that no $t_k$ might exist.} 
Recall that $\rho$ is smaller than a suitable constant.
We show that:

\begin{equation}\label{Uk-ub}
\|W(t_k)\| \leq 1- \Bigl(\frac{\rho}{2}\Bigr)^k.
\end{equation}

\noindent
The smallest interval $[a,b]$ 
defining $\|W(t_k)\|$ is in $[0,1]$.
By symmetry, we can always assume that $a+b \geq 1$.
Because $\|W(t_1)\|=0$, we can also safely
assume by induction that~(\ref{Uk-ub}) holds up to $t_k$; hence
$a\geq \frac{1}{2} (\rho/2)^k$.
Since $\|W(t)\|$ can increase only when at least one
dry agent becomes wet, ie, 
at times of the form $t=t_l$, we can 
prove~(\ref{Uk-ub}) for $t_{k+1}$ by 
showing that $\|W(t_k+1)\| \leq 1- (\rho/2)^{k+1}$.
This easily follows from 
$[0,a\rho)\cap W(t_k+1)=\emptyset$, so it suffices to prove
the latter, which we do by contradiction.
Consider an agent $i$ contributing to 
$W(t_k+1)$ with $x_i(t_k+1) < a\rho$. Agent $i$ is wet
at time $t_k+1$, so at least one agent 
in $N_i(t_k)$ was wet at time $t_k$ (possibly itself).
This implies that $M_{i,t_k}\geq a$ and, 
by~(\ref{gen-dyn}), 
$$
x_i(t_k+1)\geq (1-\rho)m_{i,t_k} + \rho M_{i,t_k} \geq a\rho, $$
which is impossible and proves~(\ref{Uk-ub}).

\vspace{0.5cm}
\begin{figure}[htb]
\begin{center}
\hspace{0cm}
\includegraphics[width=7cm]{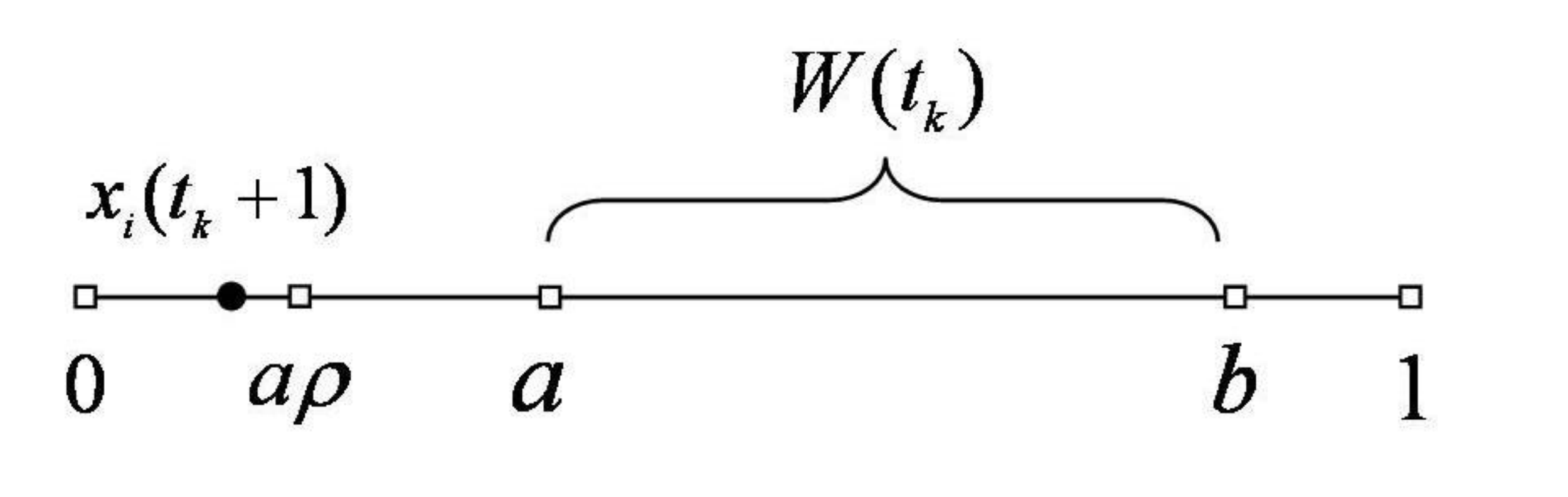}
\end{center}
\vspace{-0.0cm}
\caption{\small 
Bounding the interval spanned by wet agents.
\label{fig-recurrences-s}}
\end{figure}

\smallskip
The set $W(t_k)$ can only gain agents, as $k$ grows,
but the set may stop growing before it absorbs all of them.
When $t$ is not of the form $t_k$, the agents of $W(t)$ interact
only among themselves, so the total $s$-energy expended 
during steps $t_{k-1}+1,\ldots, t_{k}-1$
is bounded by $E_{|W(t_k)|}(s) + E_{n-|W(t_k)|}(s)$.
At time $t=t_k$, the extra energy involved is
$$\sum_{(i,j)\in G_{t}} |x_i(t)-x_j(t)|^{s}
\leq \binom{n}{2}.$$
Using obvious monotonicity properties,
it follows that, up to the highest value of $t_k$, the 
$s$-energy is bounded by 
\begin{equation*}
\sum_{l=1}^{n-1}
   \left\{ E_l(s)+ E_{n-l}(s)+  \binom{n}{2} \right\} \\
< 2nE_{n-1}(s) + n^3 .
\end{equation*}
This includes the case where no $t_k$ exists. When it does
and reaches its highest value $t$,
if $|W(t+1)|<n$ then all
the energy has been accounted for above.
Otherwise, we must add the energy expended by
the $n$ agents past $t$. 
By~(\ref{Uk-ub}), however, at time $t+1$,
the $n$ agents fit within an interval of 
length $1- (\rho/2)^{n}$.
By the scaling (power) law of the total $s$-energy,
all we need to do is 
add $(1- (\rho/2)^{n})^s E_n(s)$ to the sum;
hence~(\ref{E_n}).

The case $n=2$ is worthy of attention
because it is easy to solve exactly.
In the worst case, the two agents
start at $0$ and $1$ and move toward each other by 
the minimum allowed distance of $\rho$. This gives
us the equation
$E_2(s)= (1-2\rho)^s E_2(s) +1 $;
hence, by~(\ref{ineqs-x}),
\begin{equation}\label{E_2}
E_2(s)=
\frac{1}{1- (1-2\rho)^s} 
\leq \frac{1}{2s\rho}
\, .
\end{equation}
We now consider the case $n>2$.
By~(\ref{ineqs-x}, \ref{E_n}),
$$
E_n(s)\leq \frac{2n E_{n-1}(s) + n^3}{s(\rho/2)^n} \, .
$$
By~(\ref{E_2}) and the monotonicity of $E_n(s)$,
we verify that 
the numerator is less than $3n^3 E_{n-1}(s)$; therefore,
for $n>2$, by~(\ref{E_2}),
\begin{equation*}
E_n(s)
< \frac{3n^3 E_{n-1}(s)}{s(\rho/2)^{n}}
\leq  s^{1-n} \rho^{-n^2 - O(1)}.
\end{equation*}
This proves the upper bound of Theorem~\ref{En(s)bound}
for $s<1$. 
\hfill $\Box$
\proofend

\medskip 

\paragraph{Proof of Theorem~\ref{matrixlb-thm}.}
Recall that ${\mathcal P}$ is the family of
$n$-by-$n$ stochastic matrices such that
any $P\in {\mathcal P}$ satisfies: 
each diagonal entry is nonzero; 
no pair $p_{ij}, p_{ji}$ contains exactly one zero; 
and each positive entry is at least $\rho$.
By~(\ref{pij-rho}), the entry $(i,j)$ of a product of $t$ such matrices
can be viewed as the position of agent $i$ after
$t$ iterations of a bidirectional system with agreement parameter $\rho$, 
initialized with all the agents at $0$,
except for $j$ positioned at $x_j(0)=1$. Referring back to the
boxed algorithm, we designate agent $j$ as the one
initially wet, with all the others dry. Let $m(t)$
be the minimum value in $W(t)$. At every time $t_k$
when $W(t)$ grows in size, the minimum $m(t)$ cannot
approach $0$ closer than $\rho m(t)$. Since 
$|\{t_k\}| < n$, either agent $i$ stays dry forever
and does not leave 0 or it joins $W(t)$ and cannot be smaller
than $\min_t m(t)$, which is at least $\rho^{n-1}$.
The lower bound proof suggests a trivial construction
that achieves the very same bound and therefore proves its optimality.  
This completes the proof of Theorem~\ref{matrixlb-thm}.
\hfill $\Box$
\proofend

\subsection{The Reversible Case}\label{reversiblecase}

Our proof of Theorem~\ref{En(s)reversible}
is based on a standard use of the Dirichlet form
and classical spectral gap 
arguments~\cite{ChungSpectralGraphTheory,landauO,montenegroT}.
Let $\pi_i= q_i/\sum_j q_j$.
We easily verify that
$\pi= (\pi_1,\ldots, \pi_n)$ is the (time-invariant)
stationary distribution of the stochastic matrix
$P=P(t)$ specified by~(\ref{reversible-pij}):
\begin{equation*}
p_{ij}=
\begin{cases}
\, 1 - ( |N_i|-1 )/q_i &\text{ if $i=j$; } \\
\, 1/q_i &\text{ if $i\neq j\in N_i$; } \\
\, 0 &\text{ else.}
\end{cases}
\end{equation*}
The argument focuses on a fixed step so we may drop $t$
to simplify the notation. Let $x=(x_1,\ldots, x_n)$
and, for $u,v \in {\mathbb R}^n$, let
$\langle u,v\rangle_\pi = \sum \pi_i u_i^T v_i$.
The dynamics is invariant under translation, so we may
move the origin to ensure that
$\langle x, {\mathbf 1}\rangle_\pi =0$.
Because $\pi$ is the stationary distribution,
this property is time-invariant; in particular,
$\langle Px, {\mathbf 1}\rangle_\pi =0$.
Because $P$ is reversible, we can decompose $x= \sum_i a_i v_i$ in
an eigenbasis $\{v_i\}$ 
for $P$ orthonormal with respect to $\langle \cdot \rangle_\pi$;
all the eigenvalues are real.
Any positive $p_{ij}$ is at least $1/q_i\geq \rho$.
Let $1=\lambda_1> \lambda_2\geq\cdots\geq \lambda_{n}\geq 2\rho -1$
be the eigenvalues of $P$, with the labeling
matching the $v_i$'s. Why the inequalities?
Briefly, the gap is strict between the two
largest eigenvalues because the graph is connected;
the smallest eigenvalue is separated from $-1$ by at least 
$2\rho$ because
$(P-\rho I)/(1 - \rho)$ is itself a reversible Markov chain
(with the same eigenvectors),
hence with real-valued spectrum in $[-1,1]$.
By Perron-Frobenius, if $\mu= \max\{\lambda_2^2, \lambda_{n}^2\}$ then,
by reversibility, $\pi_i p_{ij}= \pi_j p_{ji}$, and
\begin{equation}\label{DirichletForm}
\begin{split}
\langle x,x\rangle_\pi -
\langle Px, Px\rangle_\pi
&= \langle x, (I-P^2)x\rangle_\pi
= \sum_{i,j}a_ia_j \langle v_i, (I-P^2)v_j\rangle_\pi 
= \sum_i a_i^2(1-\lambda_i^2)\\
&\geq (1-\mu)\sum_i a_i^2 
= (1-\mu)\sum_{i,j}a_ia_j \langle v_i, v_j\rangle_\pi
= (1-\mu)\langle x, x\rangle_\pi.
\end{split}
\end{equation}
Because $P$ is reversible and any nonzero 
$\pi_i p_{ij}$ is at least $\rho/n$, it holds that,
for any vector $z$,
$$\langle z, (I-P)z\rangle_\pi= 
\frac{1}{2} \sum_{i,j} \pi_i p_{ij} (z_i-z_j)^2
\geq \frac{\rho}{2n} \sum_{(i,j)\in G_t} (z_i-z_j)^2.
$$
Set $z=v_2$. By orthonormality, 
$\langle z, z\rangle_\pi= 1$ and
$\langle z, {\mathbf 1}\rangle_\pi= 0$; 
therefore,
$z$ must contain a coordinate $z_a$ such that $|z_a|\geq 1$ 
and another one, $z_b$, of opposite sign. 
Since $G_t$ is connected, there is a simple path $L$
connecting nodes $a$ and $b$. By Cauchy-Schwarz,
\begin{equation*}
\begin{split}
1-\lambda_2
&= \langle z, (I-P)z\rangle_\pi \geq
\frac{\rho}{2n} 
\sum_{(i,j)\in L} (z_i-z_j)^2 \geq
\frac{\rho}{2n^2} 
\Bigl( \sum_{(i,j)\in L} |z_i-z_j|\Bigr)^2 \\
&\geq (\rho/2n^2) (z_a-z_b)^2
=  \rho/2n^2\, .
\end{split}
\end{equation*}
Since $\lambda_{n}+1 \geq 2\rho$,
it then follows that
$$\mu \leq \Bigl( 1- \frac{\rho}{2n^2} \Bigr)^2
\leq 1- \frac{\rho}{2n^2} \, , 
$$
and, by~(\ref{DirichletForm}),
$$
\langle Px, Px\rangle_\pi
\leq \mu \langle x,x\rangle_\pi 
\leq \Bigl( 1- \frac{\rho}{2n^2} \Bigr) \langle x, x\rangle_\pi.
$$
Let $E^D_n(L,s)$ be the maximum value of the 
(diameter-based) total $s$-energy
of an $n$-agent reversible agreement system such that
$\langle x, x\rangle_\pi = L$ at time $0$.
Since $G_t$ is connected, $q_i\geq 2$; therefore
the diameter is at most
$$2\max_i |x_i| \leq 2\sqrt{L/\min_i \pi_i }
\leq \sqrt{2L n/\rho}\, ; $$
therefore,
$$
E^D_n(L,s)\leq E^D_n( ( 1-\rho/2n^2 ) L, s) 
+ (2L n/\rho)^{s/2}.
$$
The total $s$-energy obeys the scaling law 
$E^D_n(\alpha L,s)= \alpha^{s/2} E^D_n(L, s)$.
The definition of $E^D_n(s)$ assumes unit initial diameter,
which implies that $\langle x, x\rangle_\pi \leq 1$, hence
$E^D_n(s)\leq E^D_n(1,s)$ and
$$E^D_n(s)\leq 
\frac{ (2 n/\rho)^{s/2}  }
   { 1 - ( 1- \rho/2n^2 )^{s/2} }
\leq \frac{2n}{s}\Bigl( \frac{2 n}{\rho} \Bigr)^{s/2+1}
 \, ,$$
which proves Theorem~\ref{En(s)reversible}.
This follows immediately from an 
inequality we shall use again later.
For any $0\leq a,b\leq 1$,  
\begin{equation}\label{ineqs-x}
(1-a)^b\leq 1-ab.
\end{equation}
\hfill $\Box$
\proofend


\subsection{The Lower Bounds}\label{lowerbounds}

We prove the lower bounds in Theorems~\ref{En(s)bound}
and~\ref{stoptime}.

\smallskip

\paragraph{The case $s<1$.}

We describe an algorithm ${\mathcal A}_n(a,b)$ that moves
$n$ agents initially within $[a,b]$ toward
a single point $a+(b-a)y(n)$ while producing
a total $s$-energy equal to $(b-a)^s E(n,s)$.
Clearly, $E(1,s)=0$, so assume $n>1$.
We specify ${\mathcal A}_n(0,1)$ as follows.
Place $n-1$ agents at position $0$ 
and one at position $1$. The graph $G_0$ consists
of a single edge between agent 1 at position $1$ and 
agent 2 at position $0$. 
At time $0$, agent 2 moves to position $\rho$
while agent 1 shifts to $1-\rho$.
The $n-2$ other agents stay put. Next, 
apply ${\mathcal A}_{n-1}(0,\rho)$ to the set of all agents 
except $1$. By induction, we can assume that
this brings them to position $\rho y(n-1)$. Finally, apply
${\mathcal A}_n( \rho y(n-1), 1-\rho)$ to all the agents.
The operations of
${\mathcal A}_n$ leave the center of mass invariant,
so if $y(n)$ exists it must be $1/n$. 
Here is a formal argument.
The attractor point $y(n)$ satisfies the recurrence
$$
y(n)= \rho y(n-1) + (1-\rho y(n-1)-\rho)y(n),
$$
where, for consistency, $y(1)=1$.
This implies that 
$$
\frac{1}{y(n)}= 1 + \frac{1}{y(n-1)};$$
therefore $y(n)= 1/n$, as claimed. 
The total $s$-energy $E(n,s)$ satisfies the relation:
$E(1,s)=0$; and, for $n>1$,
\begin{equation*}
\begin{split}
E(n,s)&=  \rho^s E(n-1,s) 
+ (1-\rho y(n-1) -\rho)^s E(n,s) +1 \\
&\geq 
\frac{\rho^s E(n-1,s) +1}{1- (1-2\rho)^s}
\geq 
\frac{\rho^{(n-2)s}}{(1- (1-2\rho)^s)^{n-1}} \, .
\end{split}
\end{equation*}
Since $\rho$ is small enough,
$(1-2\rho)^s\geq 1-3\rho s$ and
$
E(n,s)\geq s^{1-n} \rho^{-\Omega(n)}
$,
for any $n$ large enough, 
$s\leq s_0$, and fixed $s_0<1$.
We observe that 
Algorithm ${\mathcal A}_n$ cannot start the second recursive
call before the first one is finished, which literally
takes forever. 
This technicality is easily handled, however, 
and we skip the discussion.
This completes the proof of the lower bound
of Theorem~\ref{En(s)bound} for $s<1$.
\hfill $\Box$
\proofend

\smallskip

\paragraph{The case $s=1$.}

Suppose that each $G_t$ consists of two nodes joined by an edge.
The length of the edge can be made to shrink 
by a factor of $1-2\rho$. We show that
having $n$ agents allows us to mimic
the behavior of a 2-agent system
with $\rho$ replaced by (roughly) $\rho^n$: in other words,
contraction can be made to slow down exponentially in $n$.
Without loss of generality,
we assume that $n$ is an even integer $2m\geq 4$.
Our construction is symmetric by reflection along the $X$-axis
about the origin,
so we label the agents $-m,\ldots, -1,1,\ldots, m$ from
left to right, and restrict our discussion
to the $m$ agents with positive coordinates. (Equivalently, we could
fix one agent.)
The evolution of the system consists of phases denoted
by $\theta=0,1,2,$ etc. 
At the beginning of phase $\theta$, agent $i$ 
lies at $x_1(\theta)= (1-\rho^m)^\theta$ for $i=1$ 
and at\footnote{We deviate slightly from our usual notation by
letting the argument of $x_i(\theta)$ refer
to the phase of the construction and not the time $t$.}
$$x_i(\theta)= x_{i-1}(\theta) + \rho^{i-1}(1-\rho^m)^\theta,$$
for $2\leq i\leq m$.
As usual, we assume that $\rho>0$ is small enough.
The system includes a mirror image
of this configuration about the origin at all times.
Note that all the agents are comfortably
confined to the interval $[-2,2]$, so the 
diameter $D$ is at most $4$.

We now describe the motion at phase $\theta$ in chronological order,
beginning with agent $m$. During phase $\theta$, the first graph $G_t$
($t=\theta m$)
consists of exactly two edges: one joining $m$ and $m-1$ (with
its mirror image across $x=0$); the graph $G_{t+1}$ joins
$m-1$ with $m-2$ (and its mirror image); etc. The last
graph in phase $\theta$, $G_{t+m-1}$, follows a different pattern:
it joins the two agents indexed $1$ and $-1$.
Except for $m$, all of these agents (to right of the origin) 
are moved twice during phase $\theta$:
first to the right, then to the left.
Specifically, agent $1\leq i<m$ moves right
at time $t+m-i-1$ and left at time $t+m-i$.
We use barred symbols to denote the intermediate states, 
ie, the location after the rightward moves.
At phase $\theta$,

\begin{equation*}
\hspace{1.8cm}
G_t: \begin{cases}
\, x_{m}(\theta+1)= \alpha_m x_{m-1}(\theta) + (1-\alpha_m) x_m(\theta)
= (1-\rho^m)x_{m}(\theta); \\
\, \bar x_{m-1}(\theta)= \frac{1}{2} x_{m-1}(\theta) + \frac{1}{2} x_m(\theta)
= x_{m-1}(\theta) + \frac{1}{2}  \rho^{m-1}(1-\rho^m)^\theta 
 \, .
\end{cases}
\end{equation*}
We easily verify the identities above for 
$\alpha_m= (\rho-\rho^{m+1})/(1-\rho)$. For $i=m-1,m-2,\ldots, 2$, 
with $G_{t+m-i}$ joining agent $i-1$ and $i$, the two moves are 
specified by:

\begin{equation*}
\hspace{2cm}
G_{t+m-i}: \begin{cases}
\, x_{i}(\theta+1)= \alpha_i x_{i-1}(\theta) + (1-\alpha_i) \bar x_i(\theta)
= (1-\rho^m)x_{i}(\theta) \, ; \\
\, \bar x_{i-1}(\theta)= (1-\beta_i) x_{i-1}(\theta) + \beta_i \bar x_i(\theta)
= x_{i-1}(\theta) + \frac{1}{2}  \rho^{i-1}(1-\rho^m)^\theta  \, ,
\end{cases}
\end{equation*}
where $\beta_i= 1/(2+\rho)$ and
$$\alpha_i= \frac{\rho}{2+\rho} + \frac{2(1-\rho^i)\rho^{m-i+1}}{(1-\rho)(2+\rho)} \, .
$$
Finally, at time $t+m-1$, choosing
$\alpha_1 = (\rho+2\rho^m)/(4+2\rho)$ allows us to write

\begin{equation*}
G_{t+m-1}: 
\, x_{1}(\theta+1)= -\alpha_1 \bar x_1(\theta) + (1-\alpha_1) \bar x_1(\theta)
= (1-\rho^m)x_{1}(\theta) \, .
\end{equation*}
All the coefficients $\alpha_i$ are $\Theta(\rho)$,
so we can rescale $\rho$ by a constant factor to make
the dynamics conform to a standard one-dimensional bidirectional
agreement system with parameter $\rho$ (same with the diameter $D$).
Obviously the system converges to consensus.
In each phase $\theta$, the union of the intervals formed by
the edges of all of that phase's 
graphs $G_t$ covers $[-x_m(\theta),x_m(\theta)]$;
therefore, the total $1$-energy is at least
$$ 2 \sum_{\theta=0}^\infty x_m(\theta)=
\frac{2(1-\rho^m)}{1-\rho}\sum_{\theta=0}^\infty (1-\rho^m)^\theta
>\rho^{-m}.$$
This proves 
the lower bound of Theorem~\ref{En(s)bound} for $s=1$.
For any positive $\eps<1/2$, the length
of the edge in $G_{t+m-1}$, which is $2x_1(\theta)$,
does not fall below $\eps$ until 
$\theta$ is on the order of $ \rho^{-m} \log\frac{1}{\eps}$,
which establishes the lower bound of Theorem~\ref{stoptime}.
We note that the first agent oscillates around its initial position
by roughly $\rho/2$ until $\theta$ reaches $\rho^{-\Omega(n)}$,
so the kinetic 1-energy is, like the total 1-energy, exponential in $n$.
\hfill $\Box$
\proofend

\subsection*{Acknowledgments}

I wish to thank Ali Jadbabaie and Peter Sarnak for helpful discussions,
as well as the anonymous referees for their comments.

\vspace{2cm}



\begin{thebibliography}{99}

{\small

\bibitem{angeliB}
Angeli, D., Bliman, P.-A.  
{\em Stability of leaderless multi-agent systems: 
extension of a result by Moreau},
Mathematics of Control, Signals \& Systems 18 (2006), 293--322. 

\bibitem{blondelHOT05}
Blondel, V.D., Hendrickx, J.M., Olshevsky, A., Tsitsiklis, J.N.
{\em Convergence in multiagent coordination, consensus, and flocking},
Proc. 44th IEEE Conference on Decision and Control, Seville, Spain, 2005. 

\bibitem{blondelHT09}
Blondel, V.D., Hendrickx, J.M., Tsitsiklis, J.N.
{\em On Krause's multi-agent consensus model 
with state-dependent connectivity},
IEEE Transactions on Automatic Control 54, 11 (2009), 2586--2597.

\bibitem{caoAM}
Cao, M., Morse, A.S., Anderson, B.D.O.  
{\em Coordination of an asynchronous multi-agent system via averaging},
Proceedings 2005 IFAC Congress, 2005.

\bibitem{caoSM}
Cao, M., Spielman, D.A. Morse, A.S.  
{\em A lower bound on convergence of a distributed 
network consensus algorithm},
44th IEEE Conference on Decision and Control, 
and the European Control Conference 2005,
Seville, Spain, 2005.

\bibitem{chazFlockSODA}
Chazelle, B.
{\em Natural algorithms},
Proc. 20th Annual ACM-SIAM SODA 2009, 422--431.

\bibitem{chazFlockPaper}
Chazelle, B.
{\em The convergence of bird flocking},
arXiv:0905.4241v1, 2009. 

\bibitem{chazICS}
Chazelle, B.
{\em Analytical tools for natural algorithms},
Proc. 1st Symposium on Innovations in Computer Science 
(2010), Beijing, 32--41.



\bibitem{ChungSpectralGraphTheory}
Chung, F.R.K.
{\rm Spectral Graph Theory},
CBMS Regional Conference Series in Mathematics, Vol. 92,
Amer. Math. Soc., Providence, 1997.

\bibitem{condonhernek94}
Condon, A., Hernek, D.
{\em Random walks on colored graphs},
Random Structures and Algorithms 5 (1994), 285--303.

\bibitem{CondonL}
Condon, A., Lipton, R.J.
{\em On the complexity of space bounded interactive proofs},
Proc. 30th IEEE Symp. on Foundations of Computer Science (FOCS), 
1989, pages 462--267.

\bibitem{CuckerSmale1}
Cucker, F., Smale, S.
{\em Emergent behavior in flocks},
IEEE Trans. Automatic Control 52 (2007), 852--862.

\bibitem{earlS}
Earl, M.G., Strogatz, S.H.
{\em Synchronization in oscillator networks 
with delayed coupling: a stability criterion},
Phys. Rev. E, 67 (2003), 036204(1--4).

\bibitem{grotschelLS}
Gr\"otschel, M., Lov\'asz, L., Schrijver, A.
{\rm Geometric Algorithms and Combinatorial Optimization},
Springer-Verlag, 1988.

\bibitem{hardyR}
Hardy, G.H., Riesz, M.
{\rm The General Theory of Dirichlet's Series},
Cambridge University Press, 1915.

\bibitem{hegselmanK}
Hegselmann, R., Krause, U.
{\em Opinion dynamics and bounded confidence 
models, analysis, and simulation},
J. Artificial Societies and Social Simulation 5, 3 (2002).

\bibitem{hegselmanK2006}
Hegselmann R, Krause U.
{\em Truth and cognitive division of labor: 
first steps towards a computer aided social epistemology},
J. Artificial Societies and Social Simulation 9 (2006).

\bibitem{HendrickxB}
Hendrickx, J.M., Blondel, V.D.
{\em Convergence of different linear and non-linear 
Vicsek models}, 
Proc. 17th International Symposium on Mathematical 
Theory of Networks and Systems (MTNS2006), Kyoto (Japan), 
July 2006, 1229--1240.

\bibitem{jadbabaieLM03}
Jadbabaie, A., Lin,  J., Morse, A.S.
{\em Coordination of groups of mobile autonomous agents 
using nearest neighbor rules},
IEEE Trans. Automatic Control 48 (2003), 988--1001.

\bibitem{jadbabaieMB}
Jadbabaie, A., Motee, N., Barahona, M.
{\em On the stability of the Kuramoto model of
coupled nonlinear oscillators},
Proc. American Control Conference 5 (2004), 4296--4301.

\bibitem{krause00}
Krause, U.
{\em A discrete nonlinear and non-autonomous model of 
consensus formation},
Communications in Difference Equations (2000), 227–236. 

\bibitem{kurzR}
Kurz, S., Rambau, J.
{\em On the Hegselmann-Krause conjecture in opinion dynamics},
Journal of Difference Equations and Applications (2009).

\bibitem{landauO}
Landau, H.J., Odlyzko, A.M.
{\em Bounds for eigenvalues of certain stochastic matrices},
Linear algebra and Its Applications 38 (1981), 5--15.

\bibitem{liwang2004}
Li, S., Wang, H.
{\em Multi-agent coordination using nearest 
neighbor rules: revisiting the Vicsek model}, 
2004, arXiv:cs/0407021v2

\bibitem{liXLY}
Li, C., Xu, H., Liao, X., Yu, J.
{\em Synchronization in small-world oscillator 
networks with coupling delays},
Physica A 335 (2004), 359--364.

\bibitem{lorenz03}
Lorenz, J.
{\em Multidimensional opinion dynamics when confidence changes},
Economic Complexity, Aix-en-Provence, May 2003.

\bibitem{lorenz05}
Lorenz, J.
{\em A stabilization theorem for dynamics of continuous opinions},
Physica A: Statistical Mechanics and its Applications 355
(2005), 217--223.

\bibitem{lorenzL}
Lorenz, D., Lorenz, J.
{\em Convergence to consensus by general averaging},
in R. Bru and S. Romero-Vivo (eds),
Positive Systems (2009), LNCS Springer 91--99.

\bibitem{martinez09}
Martinez, S.
{\em A convergence result for multiagent systems subject to noise},
Proc. American Control Conference (2009), St. Louis, Missouri.

\bibitem{martinezBCF07}
Martinez, S., Bullo, F., Cort\'es, J., Frazzoli, E.
{\em On synchronous robotic networks -- Part II:
Time complexity of rendezvous and deployment algorithms},
IEEE Transactions on Automatic Control 52 (2007), 2214--2226.


\bibitem{mcgrawM}
McGraw, P.N., Menzinger, M.
{\em Analysis of nonlinear synchronization
dynamics of oscillator networks
by Laplacian spectral methods},
Phys. Rev. E 75 (2007), 027104(1--4).

\bibitem{montenegroT}
Montenegro, R., Tetali, P.
{\rm Mathematical Aspects of Mixing times in Markov Chains},
Foundations and Trends in Theoretical Computer Science 1, 3 (2006),
now Publishers Inc., 237--354.

\bibitem{Moreau2005}
Moreau, L. 
{\em Stability of multiagent systems with 
time-dependent communication links},
IEEE Transactions on Automatic Control 50 (2005), 169--182.

\bibitem{olshevskyT-06}
Olshevsky, A., Tsitsiklis, J.N.
{\em Convergence speed in distributed 
consensus and averaging},
SIAM Journal on Control and Optimization 48 (2009), 33--55. 

\bibitem{papacJ05}
Papachristodoulou, A., Jadbabaie, A.
{\em Synchonization in oscillator networks:
switching topologies and non-homogeneous delays},
Proc. IEEE CDC-ECC (2005), 5692--5697.

\bibitem{papacJ06}
Papachristodoulou, A., Jadbabaie, A.
{\em Synchonization in oscillator networks with 
heterogeneous delays, switching topologies and nonlinear dynamics},
Proc. IEEE CDC-ECC (2006), 4307--4312.

\bibitem{scardoviSS}
Scardovi, L., Sarlette, A., Sepulchre, R. 
{\em Synchronization and balancing on the N-torus},
Systems \& Control Letters 56 (2007), 335--341.

\bibitem{seneta06}
Seneta, E.
{\rm Non-Negative Matrices and Markov Chains},
Springer, 2nd ed., 2006.

\bibitem{strogatz00}
Strogatz, S.H. 
{\em From Kuramoto to Crawford: exploring the 
onset of synchronization in populations 
of coupled oscillators},
Physica D 143 (2000), 1--20.

\bibitem{triplettKM}
Triplett, B.I., Klein, D.J., Morgansen, K.A.
{\em Discrete time Kuramoto models with delay},
Proc. Networked Embedded Sensing and Control Workshop,
Lecture Notes in Control and Information Sciences, Springer, 2005.

\bibitem{tsitsiklis84}
Tsitsiklis, J.N.,
{\em Problems in decentralized decision making 
and computation}, PhD thesis, MIT, 1984.

\bibitem{tsitsiklisBA}
Tsitsiklis, J.N., Bertsekas, D.P., Athans, M.
{\em Distributed asynchronous deterministic and stochastic gradient 
optimization algorithms},
IEEE Transactions on Automatic Control 31 (1986), 803--812. 

\bibitem{vicsekCBCS95}
Vicsek, T., Czir\'{o}k,  A., Ben-Jacob,  E.,
Cohen,  I., Shochet, O. 
{\em Novel type of phase transition in a system of 
self-driven particles},
Physical Review Letters 75 (1995), 1226--1229.

\bibitem{wangSCF}
Wang, Q., Duan, Z., Chen, G., Feng, Z.
{\em Synchronization in a class of weighted
complex networks with coupling delays},
Physica A 387 (2008), 5616--5622.

\bibitem{winfree67}
Winfree, A.T.
{\em Biological rhythms and the behavior
of populations of coupled oscillators}, 
J. Theoret. Bio. 16, 1 (1967), 15--42.

\bibitem{yap00}
Yap C.
{\em Fundamental Problems of Algorithmic Algebra}, 
Oxford University Press, Inc., New York, NY, 2000.

\bibitem{zanette04}
Zanette, D.H.
{\em Propagation of small perturbations
in synchronized oscillator networks},
Europhys. Lett. 68 (2004), 356.

}
\end{thebibliography}
\end{document}